\pdfoutput=1
\documentclass[aps,pre,twocolumn,groupedaddress]{revtex4-1}
\usepackage{graphicx}
\usepackage{lipsum}  
\usepackage[utf8]{inputenc}
\usepackage{mathtools}

\usepackage{float}

\usepackage{booktabs}
\usepackage[margin=1in]{geometry} 
\usepackage[usenames,dvipsnames]{xcolor}
\usepackage{parskip} 

\usepackage{relsize}
\usepackage{amsmath}    
\usepackage{amssymb}
\usepackage{bm}

\usepackage{hyperref}
\usepackage{latexsym}
\usepackage{verbatim}
\usepackage{color}
\usepackage{xcolor}
\usepackage{tabularx}
\usepackage{makecell}
\usepackage[percent]{overpic}
\enlargethispage{\baselineskip}

\def\beq{\begin{equation}}
\def\eeq{\end{equation}}

\def\beq{\begin{equation}}                          
\def\eeq{\end{equation}}                          
\def\bea{\begin{eqnarray}}                          
\def\eea{\end{eqnarray}}

\DeclareRobustCommand{\uvec}[1]{{%
  \ifcsname uvec#1\endcsname
     \csname uvec#1\endcsname
   \else
    \bm{\hat{\mathbf{#1}}}%
   \fi
}}

\preprint{}

\begin{document}


\title{Statistical Language Competition Model with 
Dynamic Edge Weighting on a Random Network}
\author{Somyaranjan Chakra}
\email{Somyaranjan.chakra.cd.phy23@itbhu.ac.in}
\affiliation{Indian Institute of Technology (BHU) Varanasi, India 221005}

\author{Mohit Anand Madhesia}
\email{mohitanandmadhesia.rs.phy23@itbhu.ac.in}
\affiliation{Indian Institute of Technology (BHU) Varanasi, India 221005}

\author{Shradha Mishra}
\email[]{smishra.phy@itbhu.ac.in}
\affiliation{Indian Institute of Technology (BHU) Varanasi, India 221005}
\date{\today}
\begin{abstract}
{This paper presents a computational study of language competition dynamics on
Erd\H{o}s--R\'enyi random networks, extending the foundational Abrams--Strogatz model
through two novel contributions: (i) a dynamic edge-weighting mechanism that reinforces
social ties between co-minority speakers by an additive increment $\Delta$, and (ii)
a probabilistic agent-based framework governing language switching via a weighted
majority rule. Phase boundaries separating the dominance and coexistence regimes
are identified across a two-dimensional parameter space $(p, \Delta)$, where $p$ denotes the
network connectivity probability. We further characterise anomalous persistence zones within predicted
dominance regions, attributing them to the formation of isolated minority speaker
clusters. Scaling study across network sizes $N \in \{50, 100, 250, 500, 1000\}$
reveal that average cluster size decreases with $N$ and that phase boundaries diffuse
with increasing stochastic noise. Finally, we discuss extensions to a tripartite
bilingual model and heterogeneous prestige/volatility to more faithfully capture real
sociolinguistic contact scenarios.
}
\end{abstract}
\maketitle

\section{Introduction}

Language competition --- the process by which one language gains or loses speakers in
contact with another --- is a well-documented sociolinguistic phenomenon with implications
for cultural heritage, policy, and community identity \cite{weinreich2010languages, hamers2000bilinguality, fishman2001can}.
The pioneering work of \cite{abrams2003modelling} brought a rigorous quantitative lens to this
problem: using two simple parameters, the prestige (or status) of a language and the
volatility of its speakers, they demonstrated that language death arises naturally as
an emergent outcome of local social interaction, without the need for complex
individual-level assumptions.

Subsequent work has enriched in Abrams--Strogatz (AS) model in several directions.
Castello \textit{et al.} introduced bilingual agents as an intermediate state, producing
richer coexistence dynamics \cite{castello2006ordering}. Castello \textit{et al.} studied community network structures
and identified mechanisms for segregated language coexistence. More recently, the role
of heterogeneous network topology --- including small-world and scale-free graphs --- has
been shown to profoundly alter phase boundaries compared to the fully-connected
mean-field approximation \cite{castello2007anomalous,vazquez2010agent}.

Despite this progress, two gaps remain underexplored. First, standard models treat
social ties as static: every edge carries equal weight regardless of shared linguistic
identity. In reality, speakers of a minority language often preferentially strengthen
ties with fellow minority speakers as a survival response. Second, probabilistic
(stochastic) switching, where an agent's language adoption is governed by a
probability rather than a forced majority rule, remains less studied on
Erd\H{o}s--R\'enyi topologies with co-evolving edge weights.

This paper addresses both gaps. We make the following contributions:

\begin{enumerate}
  \item We introduce a dynamic edge-weighting scheme in which any edge connecting
        two minority speakers is reinforced by $\Delta$ per iteration, modelling
        the social bonding between co-minority communities.
  \item We implement and analyse a probabilistic ($p$) agent-based framework that
        incorporates this weighting into a generalised Abrams--Strogatz switching
        probability.
  \item We use Support Vector Machines (SVM) to formally identify phase boundaries in
        the $(p, \Delta)$ parameter space, separating dominance from coexistence
        regimes.
  \item We characterise \emph{anomalous persistence clusters} --- localised minority
        communities that resist global convergence even within predicted dominance
        regions --- across multiple network scales.
  \item We outline a roadmap for extending the model to bilingual agents and
        heterogeneous prestige/volatility.
\end{enumerate}

The remainder of the paper is organised as follows. Section~2 reviews the theoretical
foundations. Section~3 describes the deterministic and probabilistic model variants
and the dynamic weighting scheme. Section~4 presents the phase-space analysis.
Section~5 reports anomalous cluster findings across network scales. Section~6 outlines the discussion and future work.
Section~7 discuss the bilingual extension.
Section~8 concludes our study.

\section{Theoretical Background}

\subsection{The Abrams--Strogatz Model}

Abrams -- Strogatz (AS) modelled \cite{abrams2003modelling} a population of agents that speak one of two competing
languages, A or B. The probability that a speaker of B switches to A in a given
time step is:
\begin{equation}
  P_{B \to A} = (1-s) \cdot (\sigma_A)^{a}, \qquad
  P_{A \to B} = s \cdot (\sigma_B)^{a},
  \label{eq:as}
\end{equation}
where $s \in [0,1]$ is the relative prestige of language A (so $1-s$ is the prestige
of B), $\sigma_A$, $\sigma_B$ are the local speaker fractions of A and B respectively,
and $a > 0$ is the volatility exponent controlling the non-linearity of the social
learning process. When $a < 1$, agents are highly volatile (rapid accommodation);
when $a > 1$, agents are loyal to their language.

The model has three fixed points. Two absorbing fixed points correspond to the
dominance of A or B (extinction of the other), and a third coexistence fixed point
exists but is unstable for $a > 1$ and stable for $a < 1$ \cite{vazquez2010agent}.
The sharp transition at $a = 1$ is characteristic of complex systems exhibiting
order--disorder phase transitions.

\subsection{Extensions: Bilinguals and Social Networks}

In \cite{castello2006ordering} Castello extended the AS model by introducing a bilingual state AB as
an intermediate step: transitions from A to B (and vice versa) necessarily pass
through AB  \cite{castello2006ordering}. The resulting system has richer phase behaviour; notably, the presence
of bilingual agents shifts the critical volatility for coexistence to $a \approx 0.63$
(down from $a = 1$ in the AS model), meaning higher volatility is required to sustain
linguistic diversity \cite{vazquez2010agent,castello2006ordering}.

The role of social network topology has been equally important. 
It showed that networks with community structure allow segregated coexistence --- each
community locks into a language with bilingual agents occupying boundary nodes ---
analogous to the survival of Pennsylvania Dutch in isolated Amish communities  \cite{castello2007anomalous}. In
sparse random networks, local effects reinforce the role of prestige and reduce the
parameter space supporting coexistence.

Despite this progress, two aspects of language competition dynamics remain underexplored.
First, in both the original AS model and its bilingual and network-topology extensions, social
ties are treated as static: an edge between two agents carries the same weight throughout the
simulation, regardless of the linguistic identity of the agents it connects. Yet a substantial body
of sociolinguistic evidence suggests that minority-language speakers often respond to pressure
from a dominant language by strengthening ties within their own community, a form of social
bonding that current models do not capture. Second, while network topology has been shown to
strongly influence coexistence outcomes \cite{castello2007anomalous, vazquez2010agent}, the interaction between
topology and a co-evolving, weighted social structure has received comparatively little attention,
particularly on sparse random graphs where percolation effects are most relevant. Motivated by
these gaps, we introduce a dynamic edge-weighting mechanism in which ties between co-minority
speakers are reinforced over time, and embed this mechanism within a probabilistic, weighted
generalisation of the AS switching rule on Erd\H{o}s--R\'enyi networks. This allows us to examine
how the interplay between network connectivity and adaptive social reinforcement shapes the
boundary between linguistic dominance and coexistence, complementing the bilingual and
community-structure extensions discussed above \cite{carro2016coupled}.

\section{Model Description}

\subsection{Network Initialisation}

All simulations use an Erd\H{o}s--R\'enyi random graph $G(N, p)$ with $N$ nodes and
edge probability $p$.
Here $p$ represents the probability that any pair of nodes $(i, j)$ is connected
by an undirected edge; it therefore controls the global sparsity of the social network \cite{albert2002statistical}.

\textbf{Motivation for the Erd\H{o}s--R\'enyi topology:} We adopt $G(N,p)$ as the underlying
substrate for four reasons. First, it provides a simple, minimally-structured baseline
against which the effects of more complex topologies -- such as the community and
small-world structures studied by \cite{castello2007anomalous} and \cite{vazquez2010agent} -- can later
be compared. Second, many real social networks are sparse, in the sense that each individual
maintains a roughly constant number of meaningful social ties regardless of population size;
the Erd\H{o}s--R\'enyi construction with $p \sim O(1/N)$ naturally reproduces this sparsity
regime. Third, the model has well-understood mathematical properties -- including sharp
percolation thresholds and known degree-distribution statistics -- which allow us to
interpret phase-boundary shifts in terms of established network theory rather than
topology-specific artefacts. Fourth, and most importantly, the absence of community structure,
degree heterogeneity, or clustering in $G(N,p)$ allows us to isolate the effect of
connectivity $p$ and dynamic edge weighting $\Delta$ on language coexistence, without
confounding these effects with the topological features studied elsewhere in the literature \cite{newman2018networks, albert2002statistical, bollobas2001degree}.

Nodes are initialised with equal probability to speak language A or language B:
\begin{equation}
  \rho_A(0) = \rho_B(0) = 0.5,
  \label{eq:init_density}
\end{equation}
where $\rho_A(0)$ and $\rho_B(0)$ denote the fraction of agents speaking language A
and language B at time $t=0$, respectively. The symmetric initialisation ensures
that neither language starts with a prestige advantage.

\textbf{Note on notation:} Throughout this paper, $\rho_L$ denotes a \emph{population-level
density}: the fraction of all $N$ agents speaking language $L$ at a given time. By contrast,
$\sigma_i$ (as used in Equations~(5) and~(6) denotes
the \emph{individual language state} of a specific agent $i$ -- i.e. which language ($A$ or
$B$) agent $i$ currently speaks. Thus $\rho_A(t) = \frac{1}{N}\sum_i \delta(\sigma_i(t), A)$:
$\rho$ is an aggregate statistic computed over the population, while $\sigma$ is a per-agent
variable. This distinction becomes important once individual states become heterogeneous
under the weighted dynamics introduced below.

Each edge $(i, j)$ --- where $i$ and $j$ are two connected nodes --- is initialised
with unit weight:
\begin{equation}
  w_{ij}(0) = 1.
  \label{eq:init_weight}
\end{equation}
Here, $w_{ij}(t)$ denotes the \emph{social-tie strength} between nodes $i$ and $j$ at
time $t$: a scalar quantity representing how strongly agent $i$'s language choice is
influenced by agent $j$ (and vice versa, since the network is undirected), independent
of the mere presence or absence of the edge itself. This baseline weight of 1 reflects
equal social-tie strength at the start of the simulation, before any community bonding
has taken place. As described in Section B, $w_{ij}(t)$ evolves
over time according to the linguistic identities of $i$ and $j$, allowing the model to
capture the reinforcement of ties between co-minority speakers.

\subsection{Dynamic Edge Weighting}

The core innovation of this work is a co-evolving edge weight. At each discrete time
step $t$, after language states are updated, the weight of every edge connecting two
nodes that speak the minority language is reinforced:
\begin{equation}
  w_{ij}(t+1) = w_{ij}(t) + \Delta
  \quad \text{if } \sigma_i = \sigma_j = L_{\min},
  \label{eq:weight}
\end{equation}
where $\Delta \in [0.1,\, 0.5]$ is the reinforcement strength parameter and
$L_{\min}$ denotes the currently minority language, determined at each time step by
\begin{equation}
  L_{\min}(t) =
  \begin{cases}
    A & \text{if } \rho_A(t) < \rho_B(t), \\
    B & \text{if } \rho_B(t) < \rho_A(t),
  \end{cases}
  \label{eq:lmin}
\end{equation}
i.e.\ whichever language has the smaller population density is treated as the
minority language for the purposes of edge reinforcement at time $t$. Since
$\rho_A(t) + \rho_B(t) = 1$, $L_{\min}(t)$ is simply the language with density below
$0.5$, and may switch identity over the course of the simulation if the population
balance shifts. All other edges retain their current weight \cite{carro2016coupled, seoane2018morphospace}.

\paragraph{\textbf{Motivation for the $\Delta$ range $[0.1, 0.5]$}:}
The lower bound $\Delta = 0.1$ is chosen to ensure that edge reinforcement is
\emph{dynamically meaningful}: for $\Delta < 0.1$ the weight increment per step
is too small relative to the majority-pressure term, causing the minority community
to erode before any measurable bonding effect can stabilise it. The upper bound
$\Delta = 0.5$ guards against \emph{numerical instability}: for $\Delta > 0.5$ the
intra-minority edge weights grow so rapidly that they dominate all switching
probabilities within a few iterations, producing an artefactual ``lock-in'' of
minority speakers that is not physically motivated.  The interval $[0.1, 0.5]$
thus brackets the regime where dynamic bonding is strong enough to matter but
mild enough to remain comparable to external majority pressure throughout the
simulation horizon.

\paragraph{\textbf{Motivation for the $p$ range $[0.0025, 0.1]$}:}
Real-world social networks are \emph{sparse}: the number of meaningful social
contacts per individual is roughly constant (Dunbar's number $\sim 150$),
so $p \sim O(1/N)$ for large $N$.  For $N = 1000$, this corresponds to
$p \approx 0.001$--$0.1$.  Below $p \approx 0.0025$ the network is below
the percolation threshold and fractures into isolated components, preventing
any meaningful global dynamics; above $p \approx 0.1$ the network becomes
effectively dense and mean-field results apply, removing the topological
richness that motivates this study.  The range $[0.0025, 0.1]$ therefore
spans the sociologically relevant sparse-to-near-dense transition.

This rule captures the sociolinguistic observation that minority speakers
strengthen intra-community bonds as a survival mechanism \cite{milroy1987language}.

\begin{figure}
\centering
\includegraphics[width=0.5\textwidth]{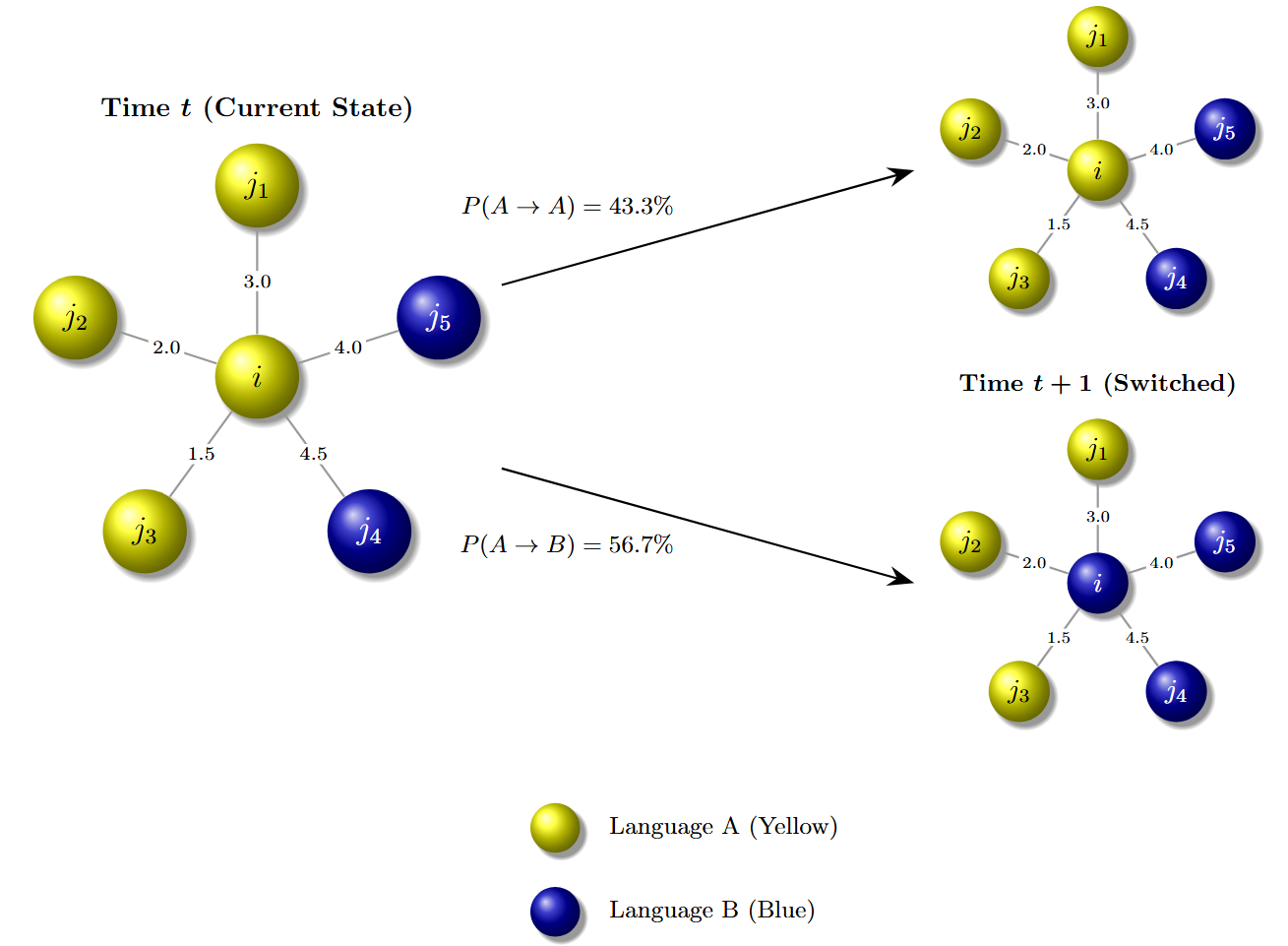}
\caption{Probabilistic Model: Schematic illustration of the probabilistic agent-based language competition model with dynamic edge weighting. Yellow nodes represent agents speaking Language~A, while blue nodes represent agents speaking Language~B. The central node updates its language state according to the weighted influence of its neighbouring agents, where the edge weights $w_{ij}$ denote the strength of social interactions. The language-switching probability is computed from the weighted local language densities using Eq.~$4$. During the evolution, edges connecting neighbouring minority-language speakers are reinforced according to Eq.~$6$, establishing a feedback between network topology and language dynamics.}
\label{fig:probabilistic_model}
\end{figure}

\subsection{Probabilistic Agent-Based Framework}

The probabilistic variant replaces the forced transition with a language-switching
probability derived from the weighted local density:
\begin{equation}
  P_i(A \to B) = s \cdot
  \left(
    \frac{\displaystyle\sum_{j \in \mathcal{N}(i),\, \sigma_j = B} w_{ij}}
         {\displaystyle\sum_{j \in \mathcal{N}(i)} w_{ij}}
  \right)^{\!a},
  \label{eq:prob}
\end{equation}
which generalises Equation $1$ to weighted, locally-connected networks.  $\mathcal{N}(i)$ is the number of connected nodes. 
The baseline simulations use $s = 0.5$ (symmetric prestige) and $a = 1$ (neutral
volatility / linear response to weighted density).  An optimisation step sets
$s = 1$ to accelerate convergence without loss of generality for the symmetric case.

Fig.$1$ provides a schematic illustration of the probabilistic
agent-based framework employed in this study. The central node $i$ interacts with its
neighbouring agents through weighted edges, where each edge weight represents the
strength of social interaction between two connected individuals. The neighbouring
nodes belong to either Language A or Language B, and the weighted contributions from
each language are accumulated separately to determine the local linguistic influence on
the central node \cite{carro2016coupled}.

The figure further demonstrates that the decision of an individual depends not only
on the number of neighbouring speakers but also on the strengths of the social ties
connecting them. Neighbours connected through larger edge weights exert a stronger
influence on the language-switching probability than those connected through weaker
links. During the evolution of the system, these interaction strengths are continuously
updated through the dynamic edge-weight reinforcement mechanism described in
Eq. $4$. This adaptive reinforcement establishes a feedback mechanism
between the evolving network structure and the language dynamics.

\subsection{Simulation Protocol}

For each parameter pair $(p, \Delta)$ drawn from the grid
$p \in [0.0025, 0.1]$, $\Delta \in [0.1, 0.5]$, the system evolves for
$T = 250$ time steps. Simulations are performed for two representative values of the language prestige parameter, $s=0.5$ and $s=1.0$, in order to examine the influence of prestige on the competition dynamics. Each time step is sufficient time in which agents evolve to a new state typically of the order of few months to a year for typical social network.  The final state is recorded as the A-language population, in a total population of $N = 1000$. Multiple realisations are averaged to reduce stochastic variance.

\begin{figure}[t]
\centering

\begin{overpic}[width=0.36\textwidth]{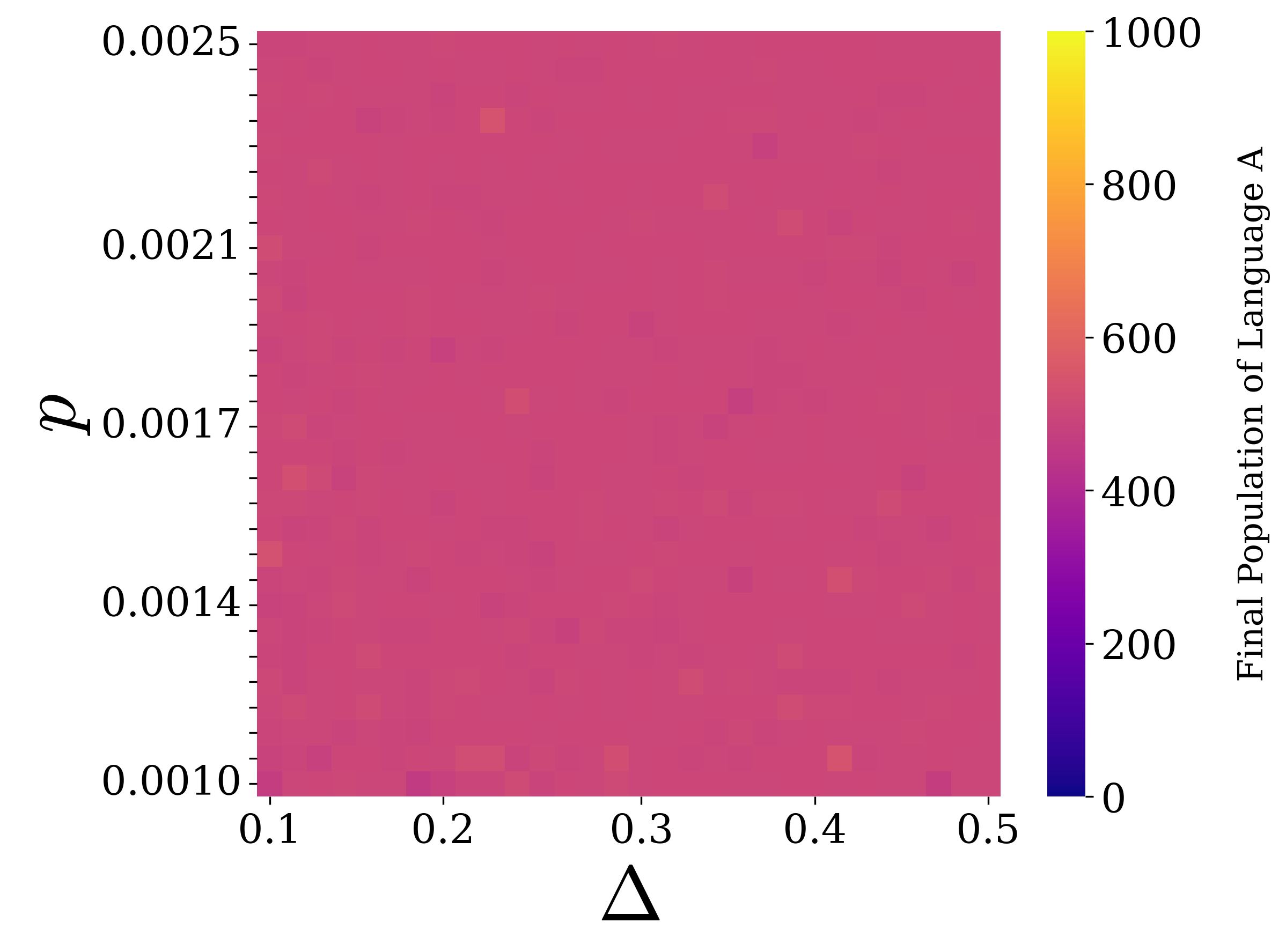}
    \put(94,68){\bfseries (a)}
\end{overpic}
\hfill
\begin{overpic}[width=0.36\textwidth]{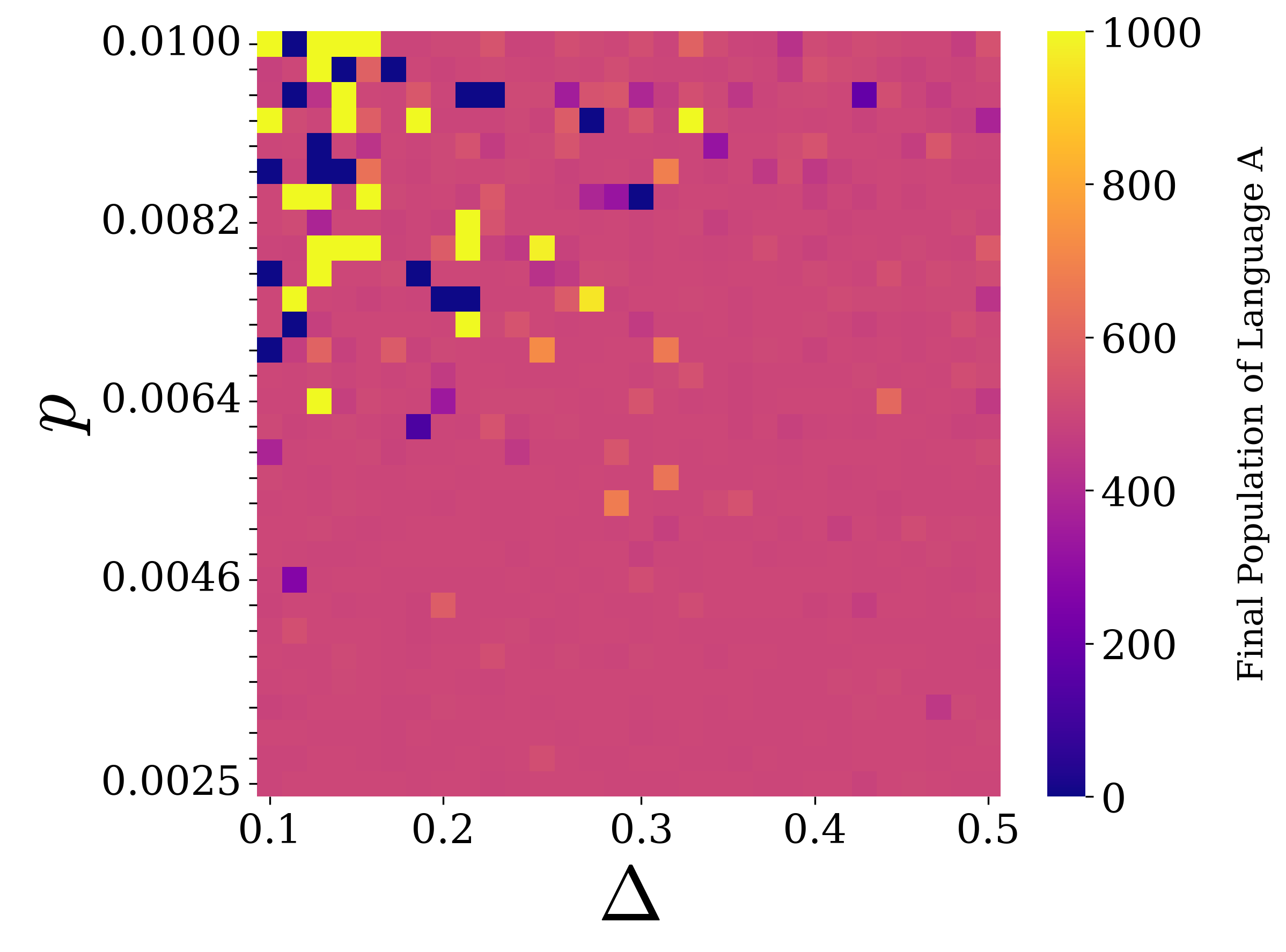}
    \put(94,68){\bfseries (b)}
\end{overpic}
\hfill
\begin{overpic}[width=0.36\textwidth]{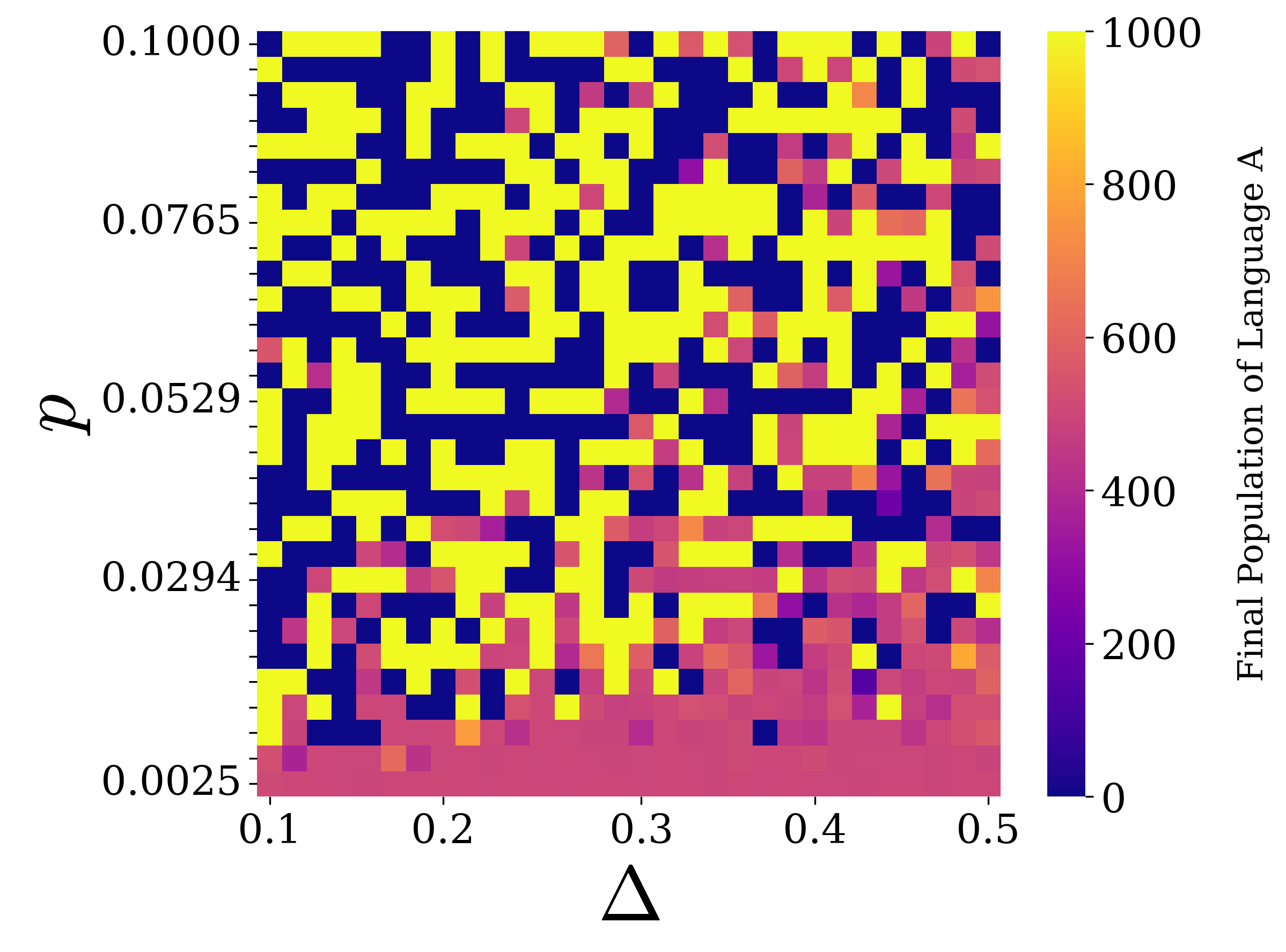}
    \put(94,68){\bfseries (c)}
\end{overpic}

\caption{Comparative phase-space diagrams for three connectivity regimes.
Colour encodes the A-language population  at $T = 250$.
(a) Sparse ($p$ low); (b) Transition (percolation threshold);
(c) Dense ($p$ high).}
\label{fig:phase_compare}
\end{figure}

\section{Phase-Space Analysis}

\subsection{Comparative Phase-Space Regimes}

We classify the final state of each simulation into two regimes based on the final
density $\rho$ of the majority language:

\begin{itemize}

  \item \textbf{Binary Dominance} ($\rho \geq 0.90$): one language captures a
        supermajority, indicating a ``winner-takes-all'' outcome.
  \item \textbf{Non-Binary / Coexistence} ($\rho < 0.90$): neither language
        dominates; linguistic diversity persists.
\end{itemize}

Three connectivity regimes are examined while sweeping $\Delta \in [0.1, 0.5]$:

Fig.$2$ presents the comparative phase-space diagrams for three different connectivity regimes of the Erd\H{o}s--R\'enyi network. The horizontal axis represents the network connectivity probability $p$, while the vertical axis denotes the dynamic edge-reinforcement parameter $\Delta$. The colour scale indicates the final majority-language density $\rho$ after $T=250$ iterations, thereby distinguishing regions of binary dominance from those exhibiting long-term coexistence.

The principal effect of increasing connectivity is a progressive shift of the coexistence region towards higher reinforcement strengths. In sparse networks, limited communication pathways restrict the spread of the majority language, allowing minority communities to survive even under relatively weak reinforcement. As the network approaches the percolation threshold, long-range communication becomes increasingly effective, intensifying the competition between majority-language propagation and adaptive minority reinforcement. In the dense regime, multiple communication paths facilitate rapid information spreading throughout the network, making minority-language persistence possible only when sufficiently strong edge reinforcement compensates for the increased majority influence.

These results demonstrate that network connectivity and adaptive edge reinforcement act as competing mechanisms that collectively determine the phase behaviour of the system. Increasing $p$ enhances global communication and promotes language homogenisation, whereas increasing $\Delta$ strengthens local minority cohesion and stabilises coexistence. The systematic displacement of the coexistence region with increasing connectivity provides the physical motivation for the quantitative phase-boundary analysis presented in the following section.

Overall, Fig.$2$ demonstrates that network connectivity is one of the key factors governing the language competition dynamics. As the connectivity probability $p$ increases from the sparse to the dense regime, the coexistence region progressively shrinks and shifts towards larger values of $\Delta$. This behaviour indicates that increasing network connectivity enhances the efficiency of majority-language propagation, whereas dynamic edge reinforcement acts as a stabilising mechanism that enables minority-language communities to survive. These observations provide the physical basis for the quantitative phase-boundary analysis performed using the Support Vector Machine (SVM) classifier in the following section.

\begin{table}[H]
\centering
\small
\caption{Connectivity regimes and qualitative outcomes.}
\begin{tabular}{lll}
\toprule
Regime & $p$ range & Qualitative observation \\
\midrule

Sparse &
$[0.0025,0.005]$ &
Disconnected clusters;\\
& &
minority language persists\\
& &
at low $\Delta$. \\

Transition & 
$[0.005, 0.0075]$  & 
Percolation threshold;\\
& & 
boundary shifts; moderate\\
& &
$\Delta$ needed. \\

Dense      & 
$[0.0075, 0.1]$   &
Well-connected; minority\\
& &
language highly vulnerable;\\
& & 
high $\Delta$ required. \\
\bottomrule
\end{tabular}
\end{table}

Table I. summarises the qualitative behaviour of the proposed language competition model across the three connectivity regimes of the Erd\H{o}s--R\'enyi network. The results demonstrate that network connectivity plays a decisive role in determining whether the minority language survives or is driven to extinction.

In the sparse regime ($p \in [0.0025,\,0.005]$), the network is fragmented into several disconnected or weakly connected components. The limited number of communication paths reduces the ability of the majority language to propagate throughout the system. Consequently, minority-language speakers remain confined within isolated local communities, allowing them to survive even when the edge-reinforcement strength $\Delta$ is relatively small. As a result, the coexistence region occupies a larger portion of the phase space as shown in Fig.2(a)

As in Fig.2(b) the connectivity increases into the transition regime ($p \in [0.005,\,0.0075]$), the network approaches the percolation threshold and isolated components begin to merge into a giant connected cluster. At this stage, linguistic influence propagates over much larger distances, increasing the competition between majority-language expansion and minority-language reinforcement. Consequently, a moderate value of the reinforcement parameter $\Delta$ is required to maintain stable coexistence, and the phase boundary becomes increasingly sensitive to small variations in the model parameters.

In the dense regime Fig.2(c) ($p \in [0.0075,\,0.1]$), the network is highly interconnected, allowing the majority language to spread efficiently through multiple interaction pathways. The increased global connectivity significantly weakens the ability of isolated minority-language communities to resist majority influence. Therefore, only sufficiently large reinforcement strengths can preserve minority-language clusters, whereas smaller values of $\Delta$ rapidly lead to complete language dominance.

Overall, Table $1$ highlights the competing roles of network connectivity and adaptive edge reinforcement. Increasing the connectivity probability enhances the propagation of the majority language, whereas dynamic edge reinforcement strengthens interactions among minority-language speakers and promotes their persistence. The balance between these two mechanisms determines the location of the phase boundary separating dominance from coexistence and provides the physical basis for the quantitative Support Vector Machine (SVM) analysis presented in the following section.

The key qualitative finding is that increasing $p$ facilitates global communication,
exposing minority speakers to stronger majority influence and making the minority
language more vulnerable unless balanced by a sufficiently large reinforcement
factor $\Delta$.

\subsection{SVM Phase Boundary Identification}

To mark the phase boundary, we train a Support Vector Machine (SVM) \cite{joachims2002support,cortes1995support,scholkopf2002learning}
with a Radial Basis Function (RBF) kernel \cite{schaback2007practical} on the $(p, \Delta)$ parameter grid.
Each grid point is labelled $y \in \{-1, 1\}$ based on the dominance threshold.

\paragraph{Background on SVMs.}
A Support Vector Machine is a supervised binary classifier that finds the
maximum-margin hyperplane separating two classes in a (possibly
high-dimensional) feature space. Each training pair $(\mathbf{x}_i, y_i)$,
for $i = 1, \dots, n$, consists of a $(p,\Delta)$ grid point $\mathbf{x}_i$
and its regime label $y_i \in \{-1,+1\}$, where $+1$ denotes Dominance and
$-1$ denotes Coexistence.

We adopt the \emph{soft-margin} formulation, which tolerates noisy or
overlapping grid points near the transition rather than demanding perfect
separability:
\begin{equation}
  y_i\left(\mathbf{w}^\top\mathbf{x}_i + b\right) \geq 1 - \xi_i,
  \label{eq:softmargin}
\end{equation}
where $\mathbf{w}$ is the weight vector orienting the separating hyperplane
(its magnitude sets the margin width, $\propto 1/\|\mathbf{w}\|$), $b$ is the
offset, and $\xi_i \geq 0$ is a slack variable — the penalty incurred if
point $i$ falls inside the margin or on the wrong side of it. The signed
quantity $y_i\big(\mathbf{w}^\top\phi(\mathbf{x}_i)+b\big)$ is simply a
classification score: positive means $\mathbf{x}_i$ is correctly assigned to
its regime, negative means misclassified.

Training solves the convex quadratic programme
\begin{equation}
  \min_{\mathbf{w},b,\boldsymbol{\xi}}\;
  \frac{1}{2}\|\mathbf{w}\|^{2} + C\sum_{i=1}^{n}\xi_i,
  \label{eq:svm}
\end{equation}
subject to Eq.~\eqref{eq:softmargin}. The first term maximises the margin
width; the second, scaled by the regularisation hyperparameter $C > 0$,
penalises margin violations across the grid. $C$ thus controls the
trade-off between a wide, robust margin and strict classification accuracy
on the $(p,\Delta)$ data — a larger $C$ pushes the classifier to fit the
simulated dominance/coexistence labels more tightly, at the cost of a
narrower margin.

Rather than working explicitly in a high-dimensional feature space
$\phi(\mathbf{x})$, we exploit the \emph{kernel trick}: both training and
the resulting decision function depend on the data only through inner
products $\phi(\mathbf{x}_i)^\top\phi(\mathbf{x}_j)$, evaluated directly via
a kernel $K(\mathbf{x}_i,\mathbf{x}_j)$ without ever computing $\phi(\cdot)$.
The \emph{RBF kernel}
\begin{equation}
  K(\mathbf{x},\mathbf{x}') = \exp\!\left(-\gamma\|\mathbf{x}-\mathbf{x}'\|^2\right)
\end{equation}
is used here because it captures arbitrarily non-linear phase boundaries in
the $(p,\Delta)$ plane without an explicit parametric form
\cite{scholkopf2002learning,bishop2006pattern}. The resulting decision
boundary traces the critical reinforcement threshold $\Delta_c(p)$ below
which dominance occurs.

As $p$ increases across the three regime+s, $\Delta_c$ shifts upward: denser networks
require stronger minority bonding to sustain coexistence.

\begin{figure}[t]
\centering

\begin{overpic}[width=0.32\textwidth]{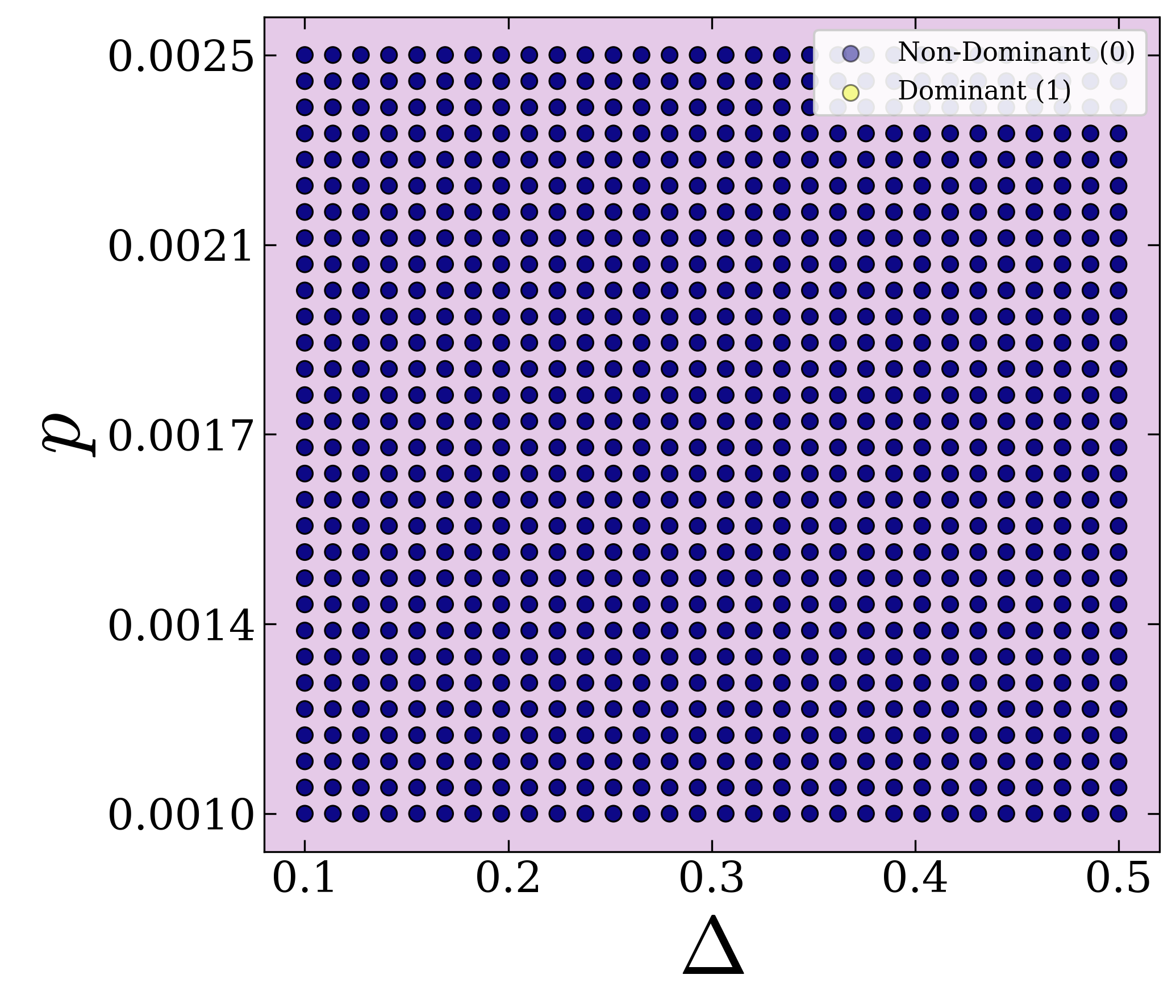}
    \put(100,76){\bfseries (a)}
\end{overpic}
\hfill
\begin{overpic}[width=0.32\textwidth]{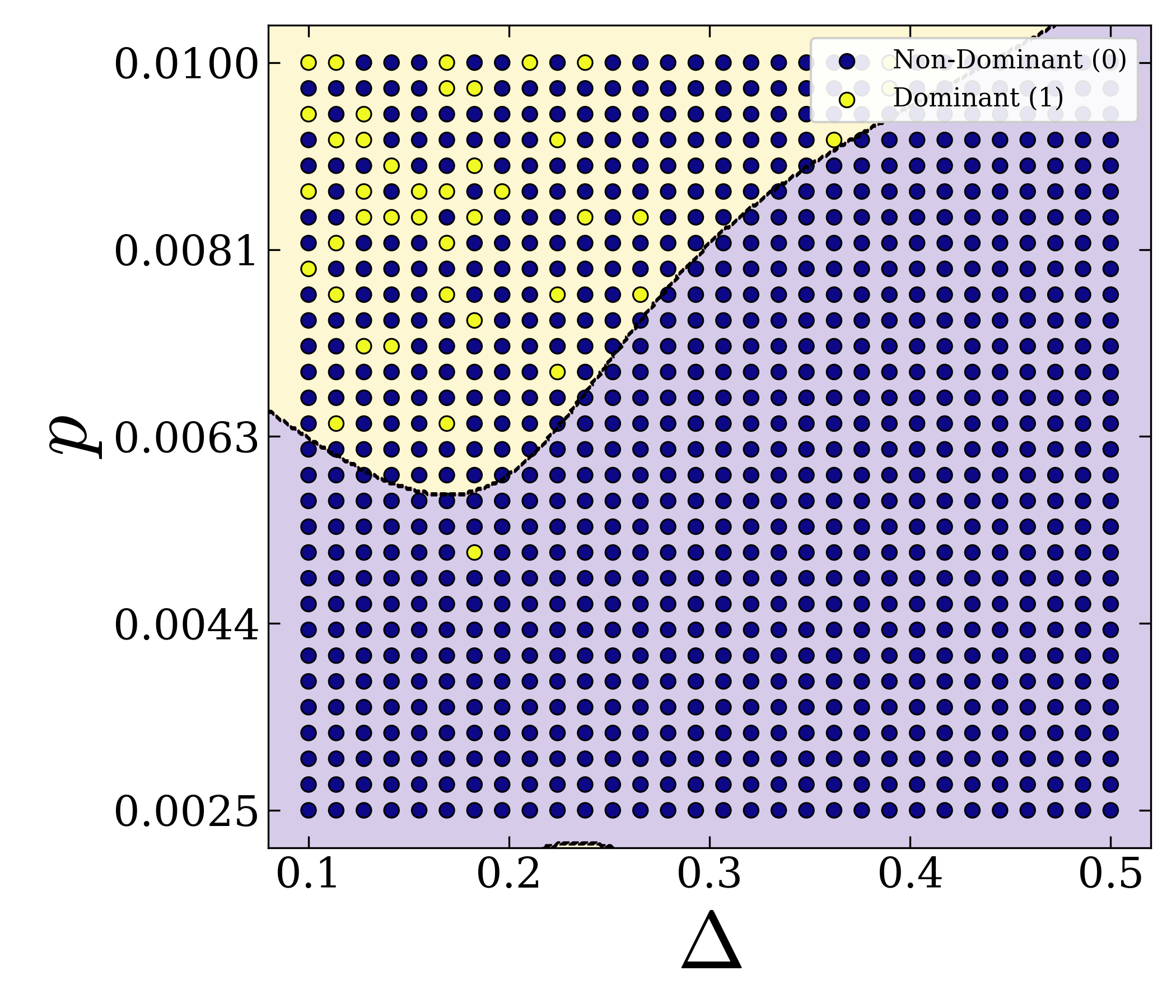}
    \put(100,76){\bfseries (b)}
\end{overpic}
\hfill
\begin{overpic}[width=0.32\textwidth]{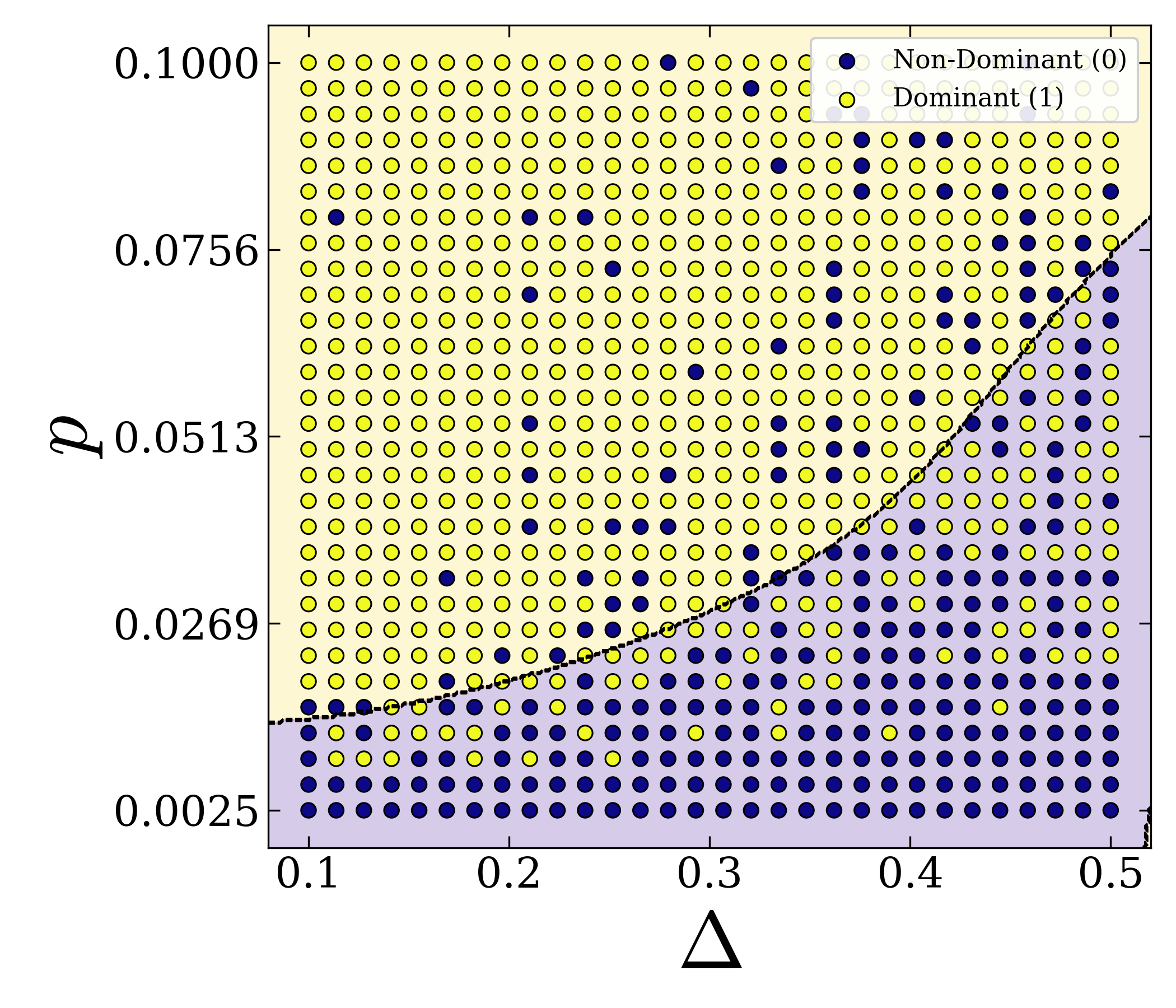}
    \put(100,76){\bfseries (c)}
\end{overpic}

\caption{SVM decision boundaries (solid curves) separating dominance from
coexistence regimes across three connectivity ranges.
(a)~Regime~1: $p \in [0.001, 0.0025]$;
(b)~Regime~2: $p \in [0.0025, 0.01]$;
(c)~Regime~3: $p \in [0.0025, 0.1]$.
The upward shift of the boundary with increasing $p$ shows that denser
networks demand higher reinforcement $\Delta$ to maintain linguistic diversity.}
\label{fig:svm_bounds}
\end{figure}

Fig.$3$ presents the Support Vector Machine (SVM) decision boundaries separating the binary dominance and coexistence regimes in the $(p,\Delta)$ parameter space for three different connectivity ranges. The solid curves represent the critical reinforcement strength required to sustain minority-language coexistence, while the coloured regions correspond to the two distinct dynamical phases identified from the simulation data. The SVM classifier provides a smooth, non-parametric approximation to the phase boundary, enabling a quantitative determination of the transition between language dominance and coexistence.

A clear systematic evolution of the phase boundary is observed as the network connectivity increases. In the lowest connectivity range, the critical reinforcement strength remains relatively small because sparse networks limit the propagation of the majority language, enabling minority communities to survive under comparatively weak adaptive reinforcement. As the connectivity approaches the percolation regime, long-range communication becomes increasingly effective, strengthening the competition between majority-language spreading and minority-language reinforcement. This competition shifts the critical boundary towards larger values of $\Delta$ and broadens the transition region. For the highest connectivity range, the network becomes sufficiently well connected that majority-language influence propagates efficiently throughout the system, and only strong adaptive reinforcement is capable of maintaining stable minority-language communities.

The progressive upward displacement of the SVM boundary across the three connectivity ranges demonstrates that the critical reinforcement required for coexistence is strongly controlled by network topology. Increasing connectivity enhances global information transfer and therefore favours majority-language dominance, whereas adaptive edge reinforcement counteracts this tendency by strengthening interactions within minority-language communities. The balance between these competing mechanisms determines the location of the phase transition in the parameter space.

The smoothness of the SVM boundaries further indicates that the transition between the two dynamical phases is continuous over the explored parameter range. Rather than producing abrupt discontinuities, changes in connectivity gradually modify the reinforcement threshold required for minority-language survival. This result highlights the effectiveness of machine-learning-based classification in identifying complex phase boundaries in stochastic agent-based systems and provides a quantitative framework for analysing the influence of adaptive social interactions on language competition.

\begin{figure*}[t]
\centering

\includegraphics[width=0.32\textwidth]{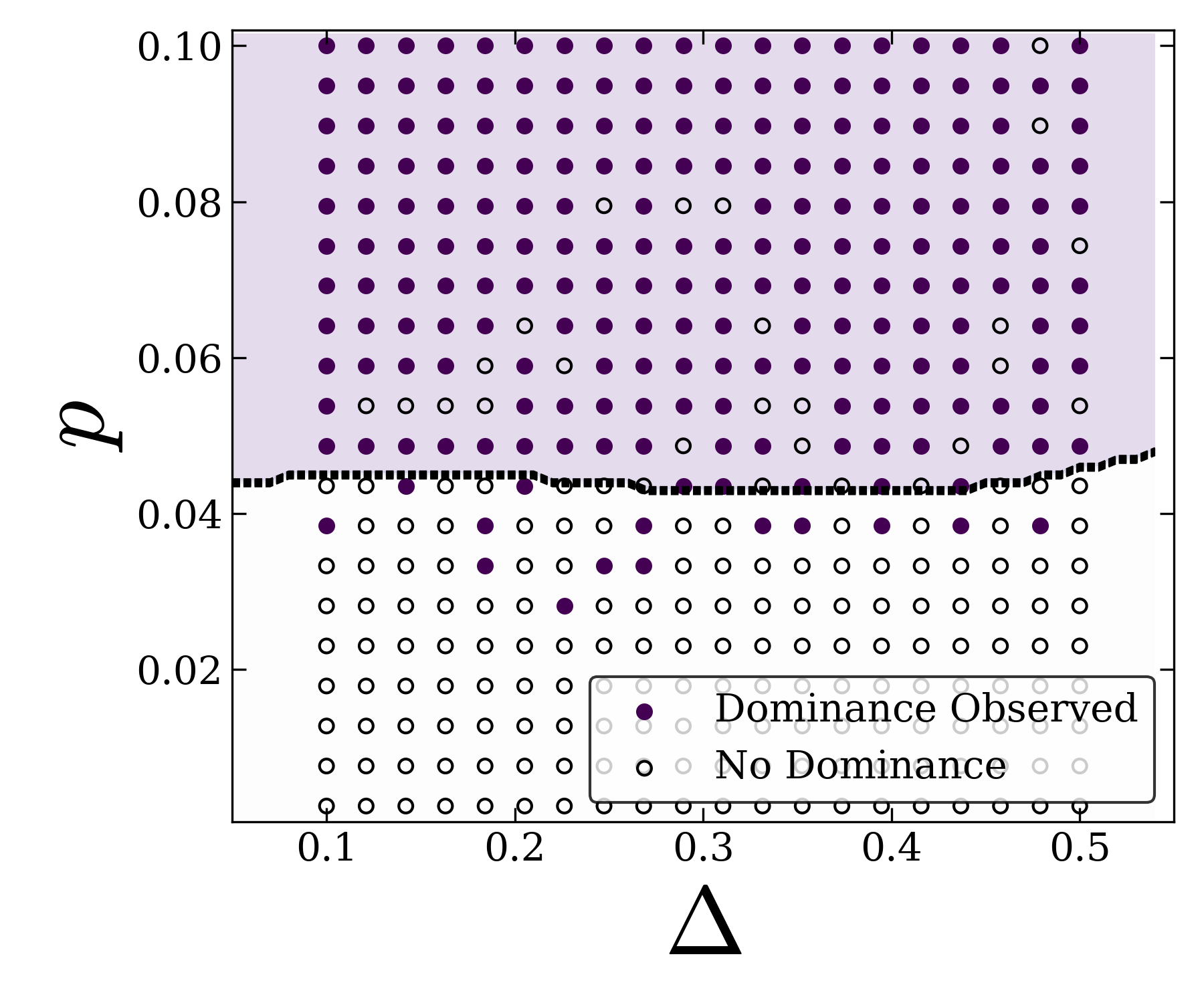}
\hfill
\includegraphics[width=0.32\textwidth]{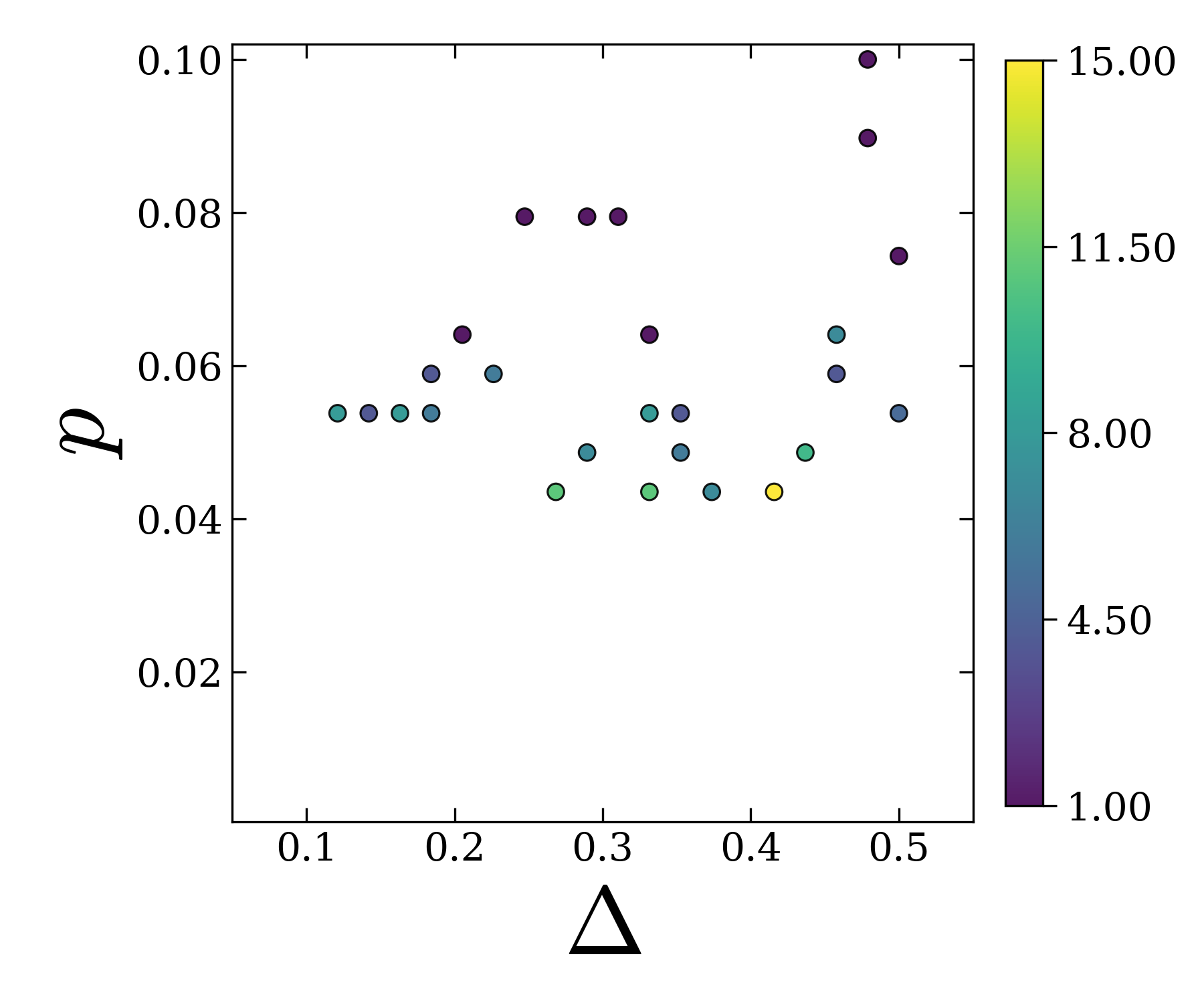}
\hfill
\includegraphics[width=0.32\textwidth]{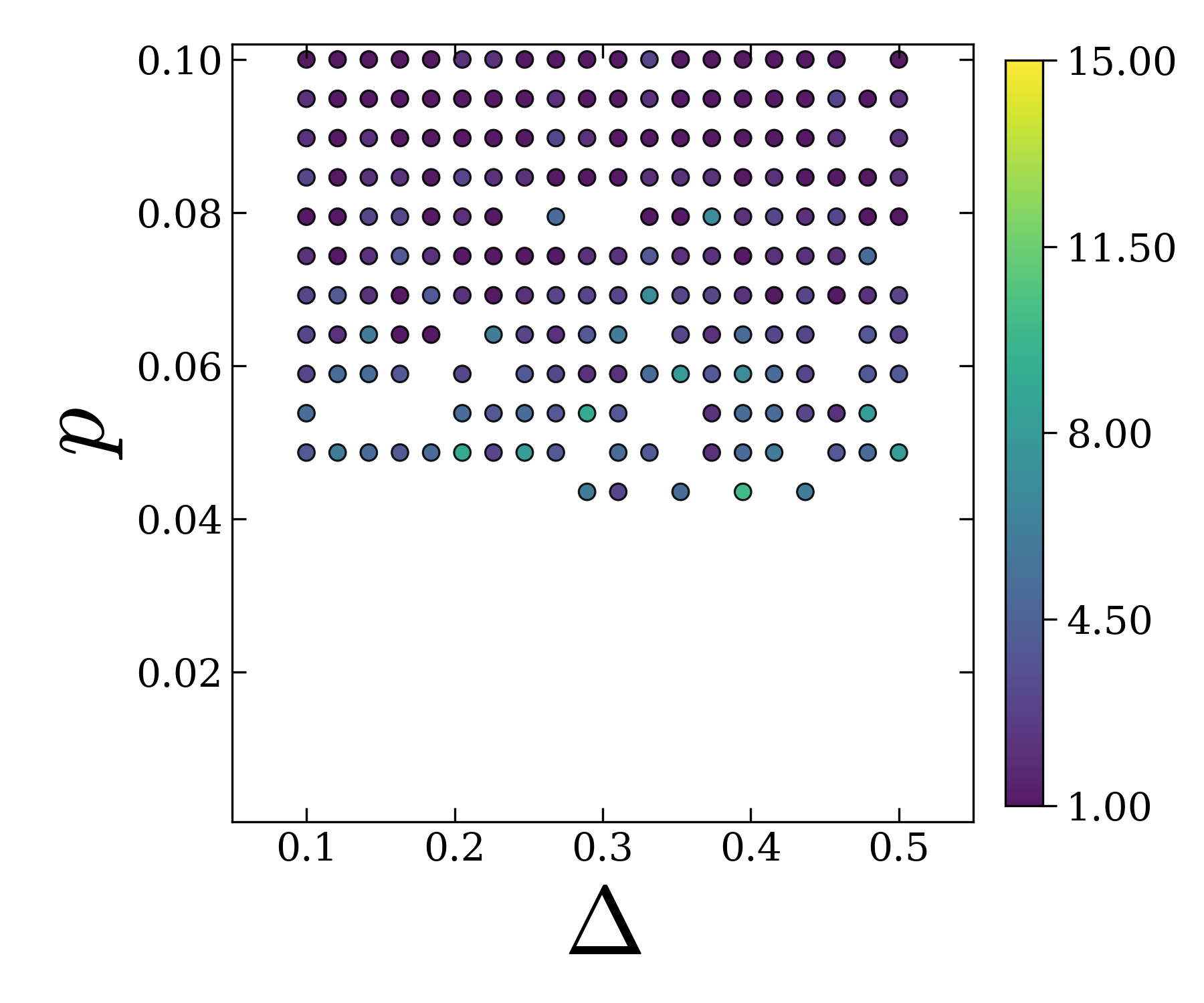}
\vspace{-0.15cm}

\makebox[0.32\textwidth]{\small (a)}
\hfill
\makebox[0.32\textwidth]{\small (b)}
\hfill
\makebox[0.32\textwidth]{\small (c)}

\vspace{0.15cm}

\includegraphics[width=0.32\textwidth]{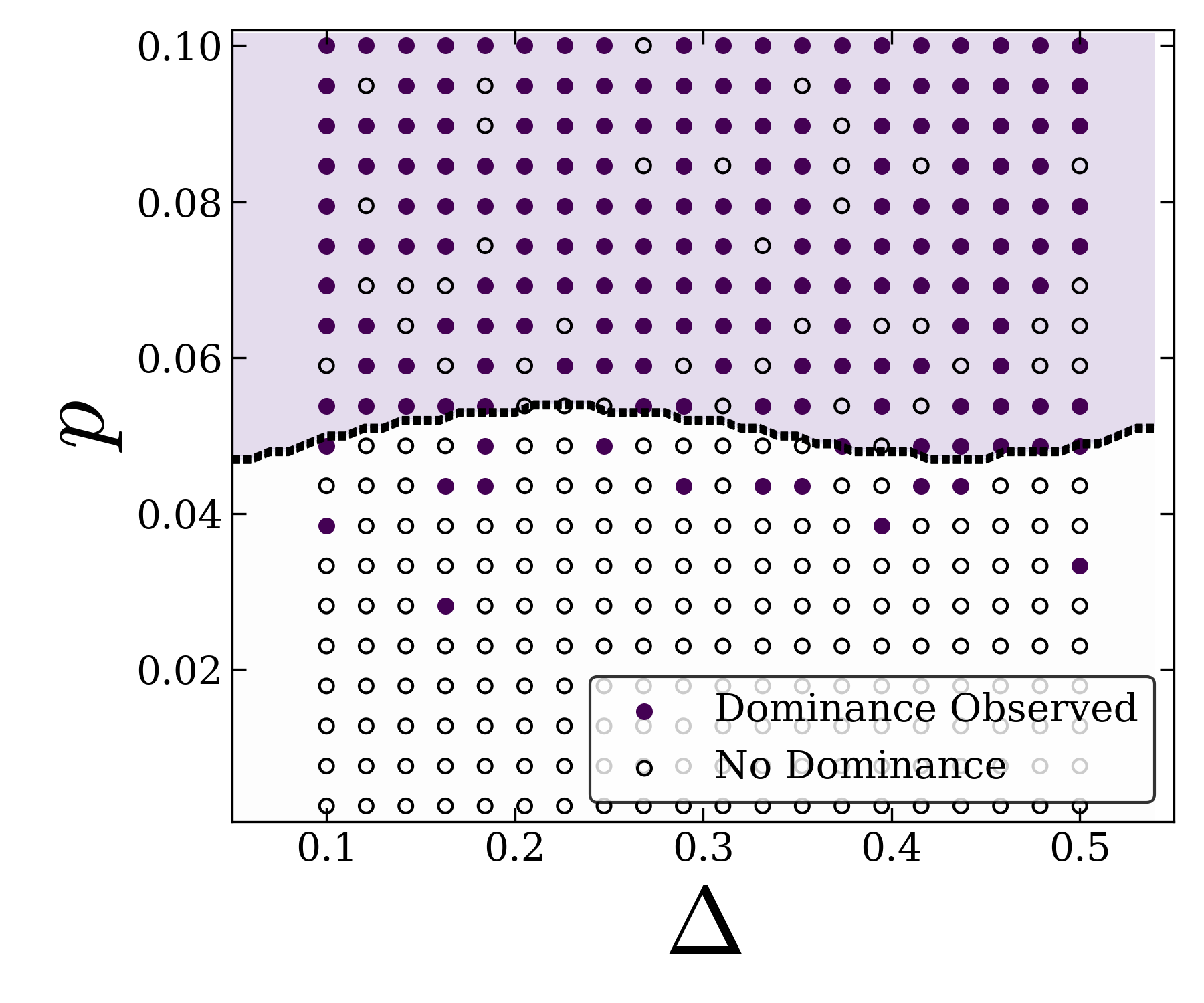}
\hfill
\includegraphics[width=0.32\textwidth]{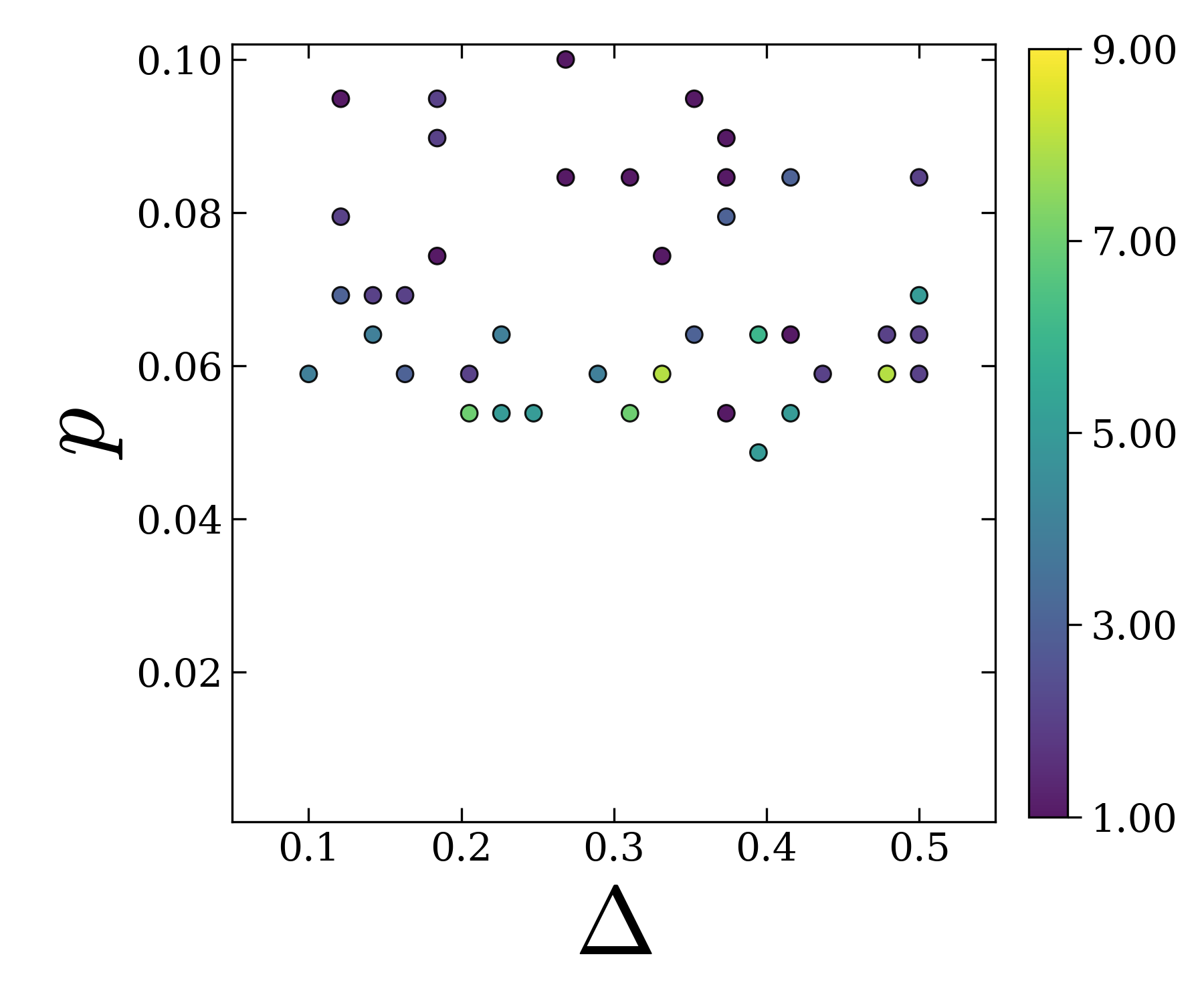}
\hfill
\includegraphics[width=0.32\textwidth]{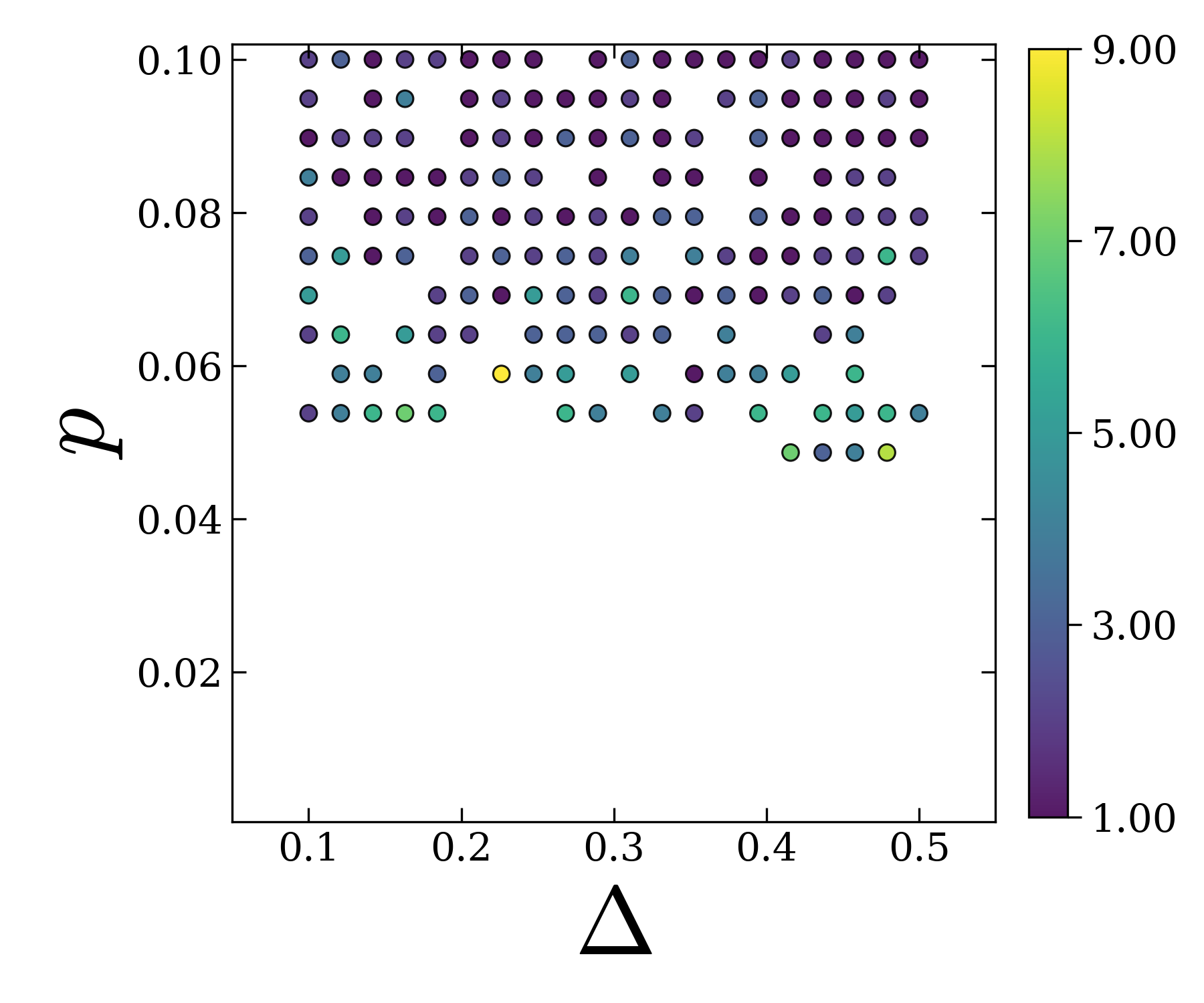}

\vspace{-0.15cm}

\makebox[0.32\textwidth]{\small (d)}
\hfill
\makebox[0.32\textwidth]{\small (e)}
\hfill
\makebox[0.32\textwidth]{\small (f)}

\caption{
Phase boundary and anomalous cluster visualisation for a network of
$N=50$ at $T=250$.
The \textbf{top row} (a-c) corresponds to $s=1.0$, while the
\textbf{bottom row} (d-f) corresponds to $s=0.5$.
Panels (a,d) show the SVM phase boundary, where open white circles denote
non-binary dominant (anomalous) points within the predicted dominance region.
Panels (b,e) show the cluster formation in anomalous points, and
panels (c,f) show the cluster formation in non-anomalous points. Colors show the size of the clusters.
}

\label{fig:anom_50}
\end{figure*}

\begin{figure*}[!htb]
\centering

\includegraphics[width=0.32\textwidth]{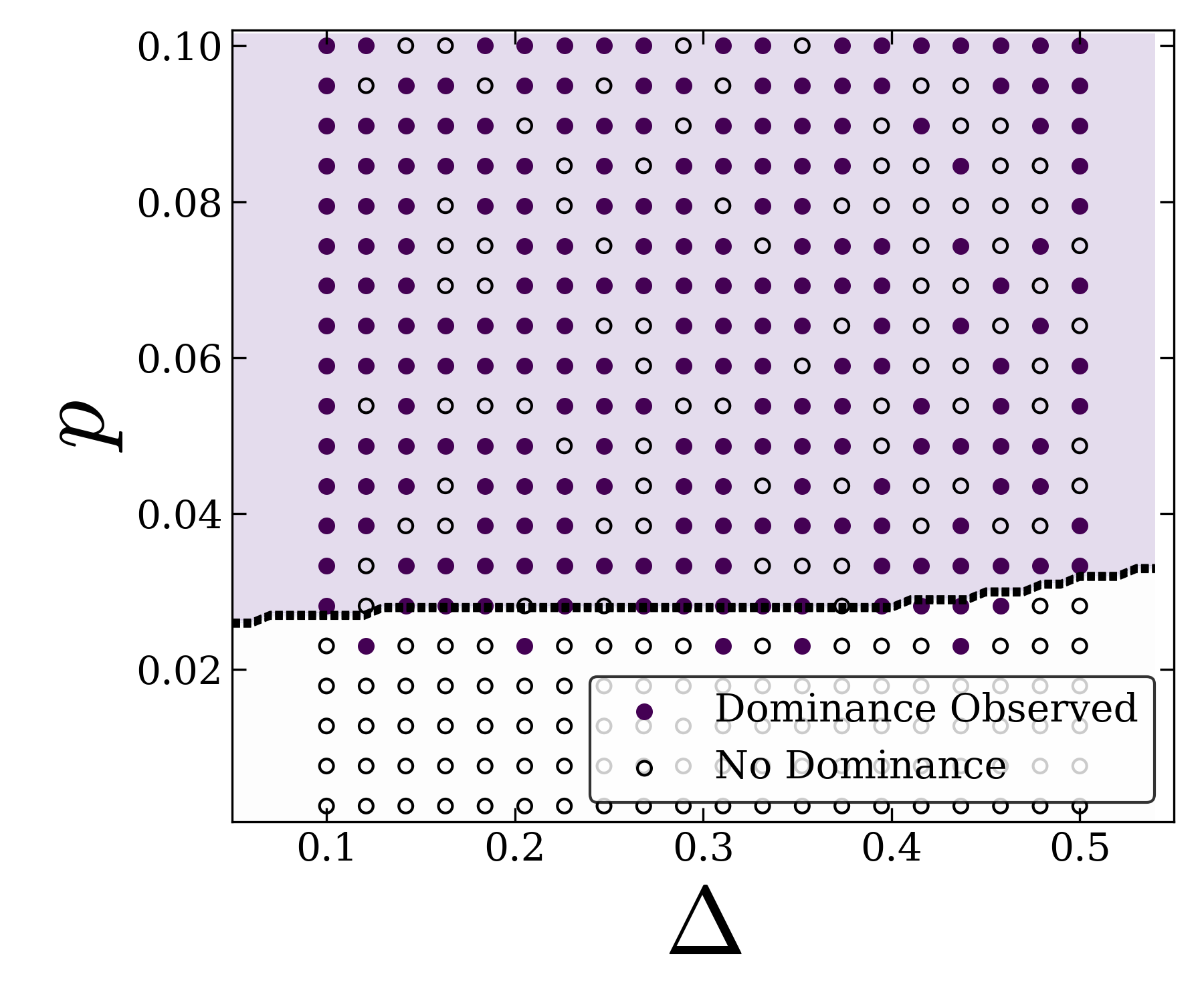}
\hfill
\includegraphics[width=0.32\textwidth]{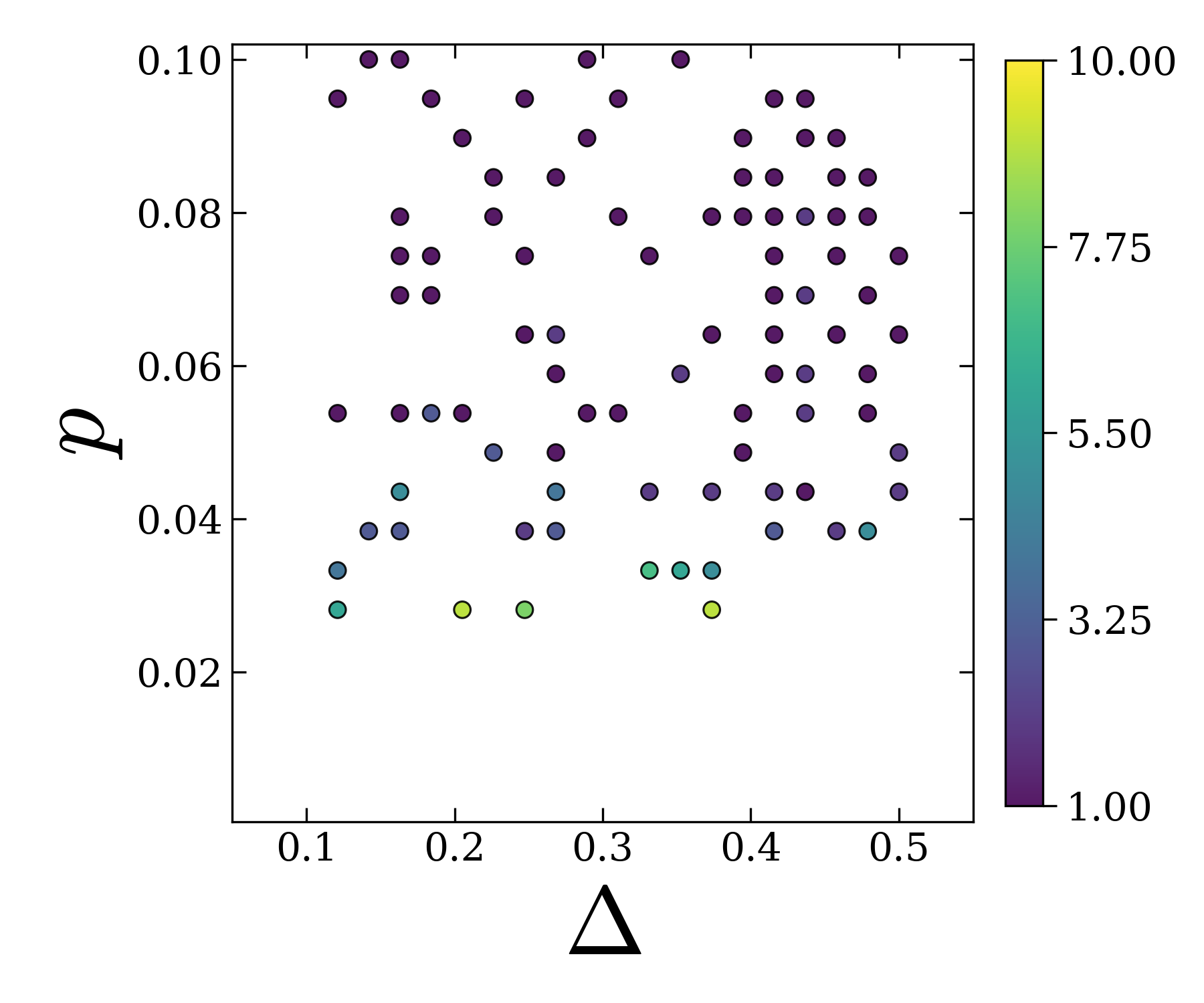}
\hfill
\includegraphics[width=0.32\textwidth]{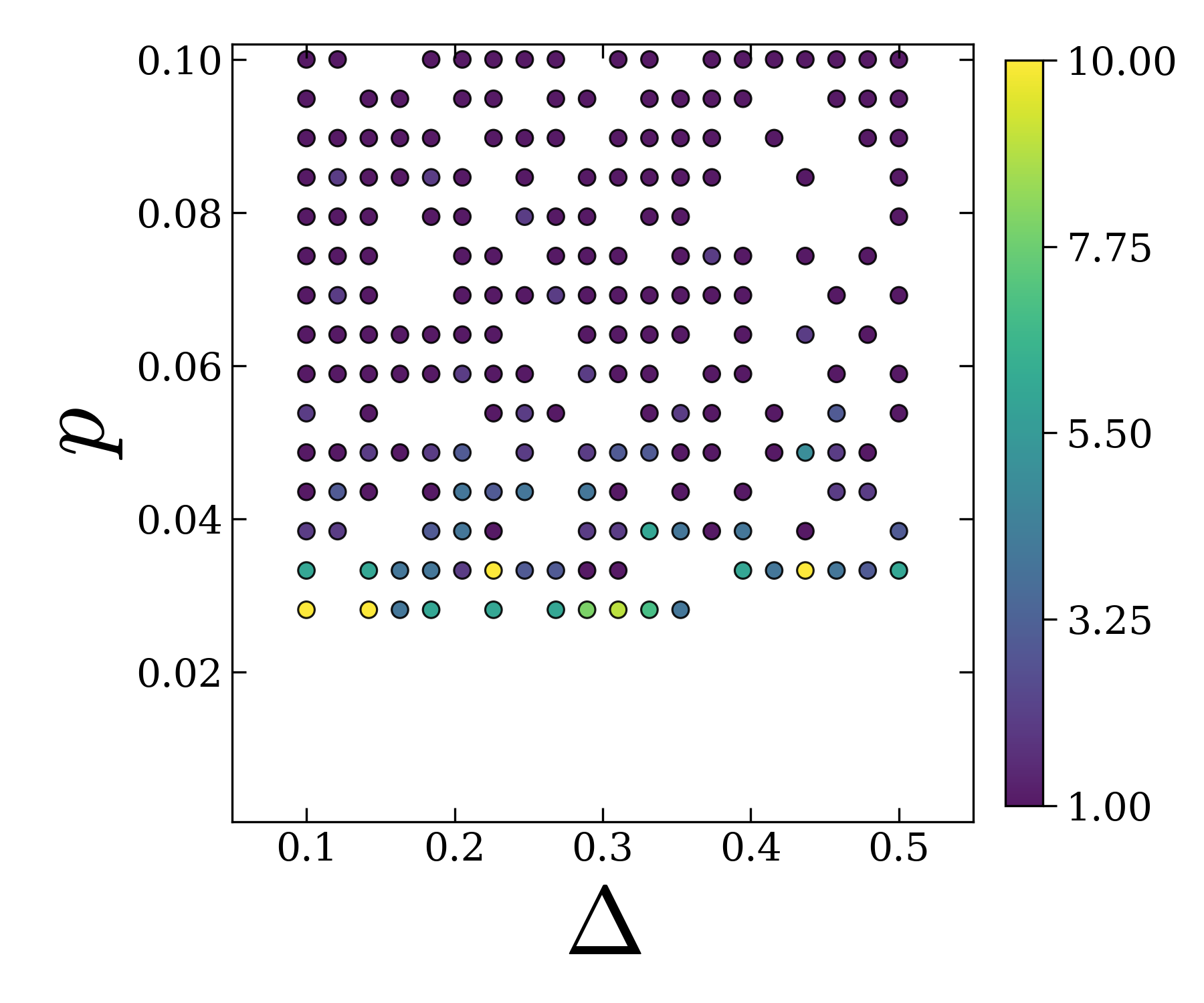}

\vspace{-0.15cm}

\makebox[0.32\textwidth]{\small (a)}
\hfill
\makebox[0.32\textwidth]{\small (b)}
\hfill
\makebox[0.32\textwidth]{\small (c)}

\vspace{0.15cm}

\includegraphics[width=0.32\textwidth]{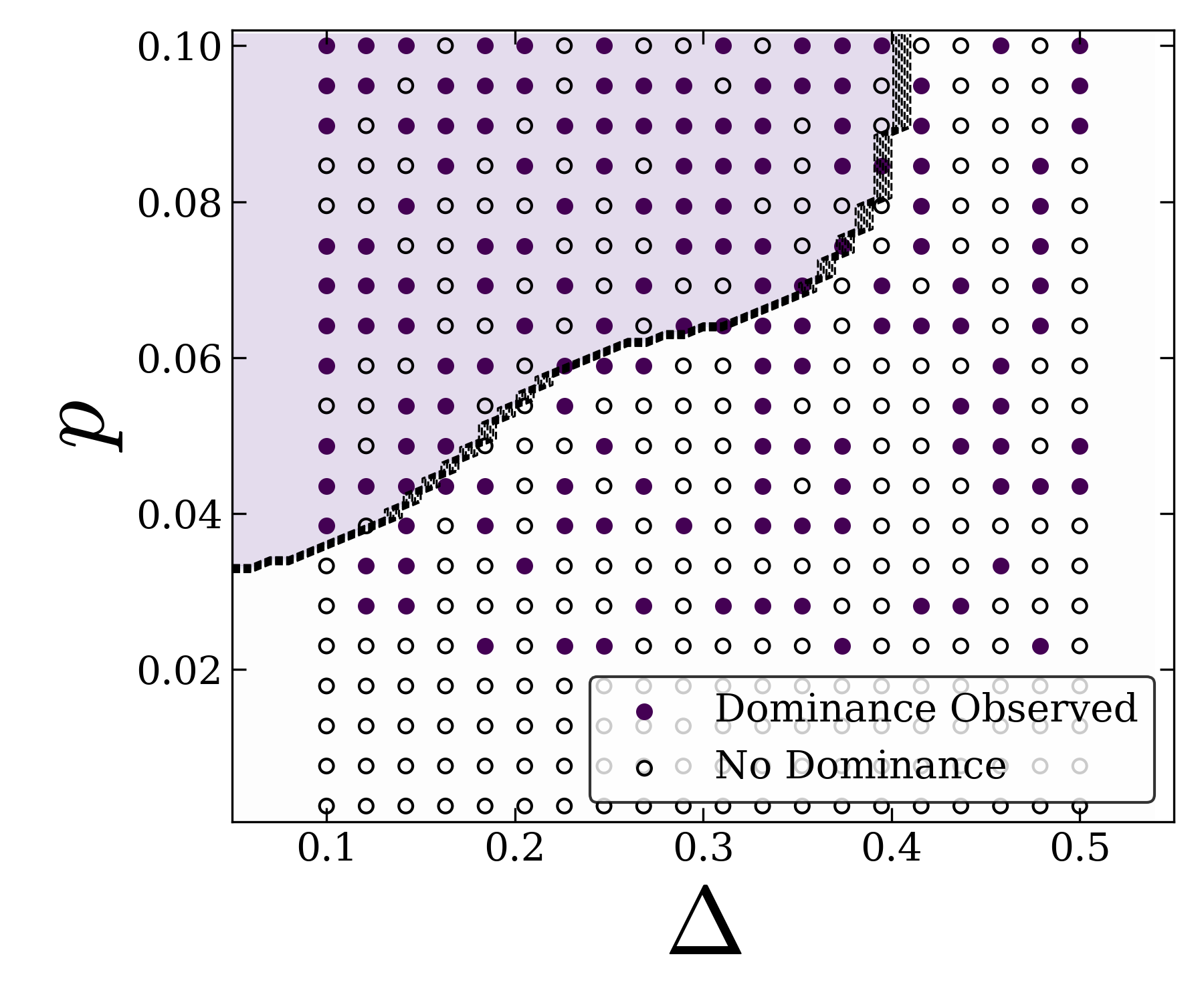}
\hfill
\includegraphics[width=0.32\textwidth]{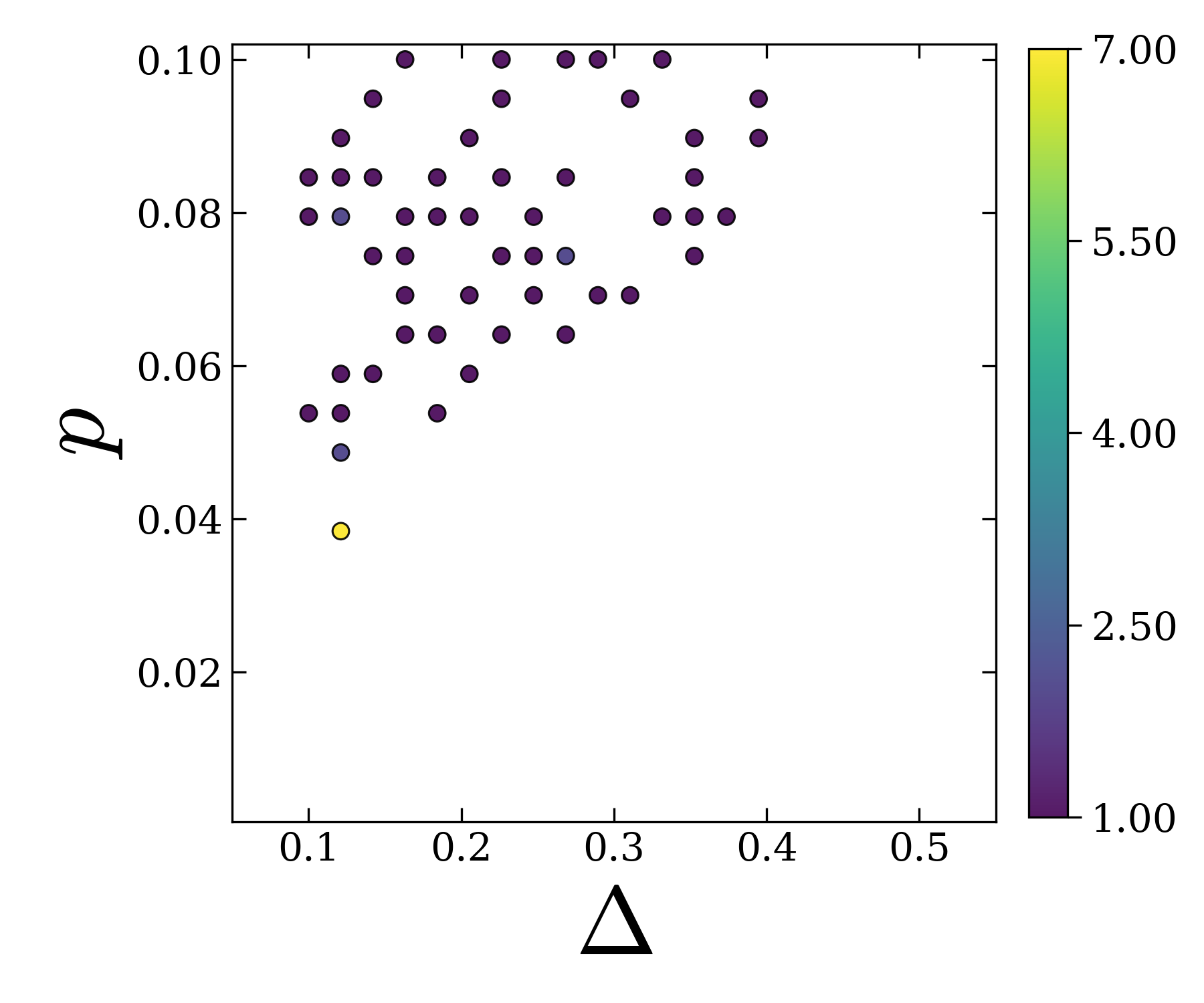}
\hfill
\includegraphics[width=0.32\textwidth]{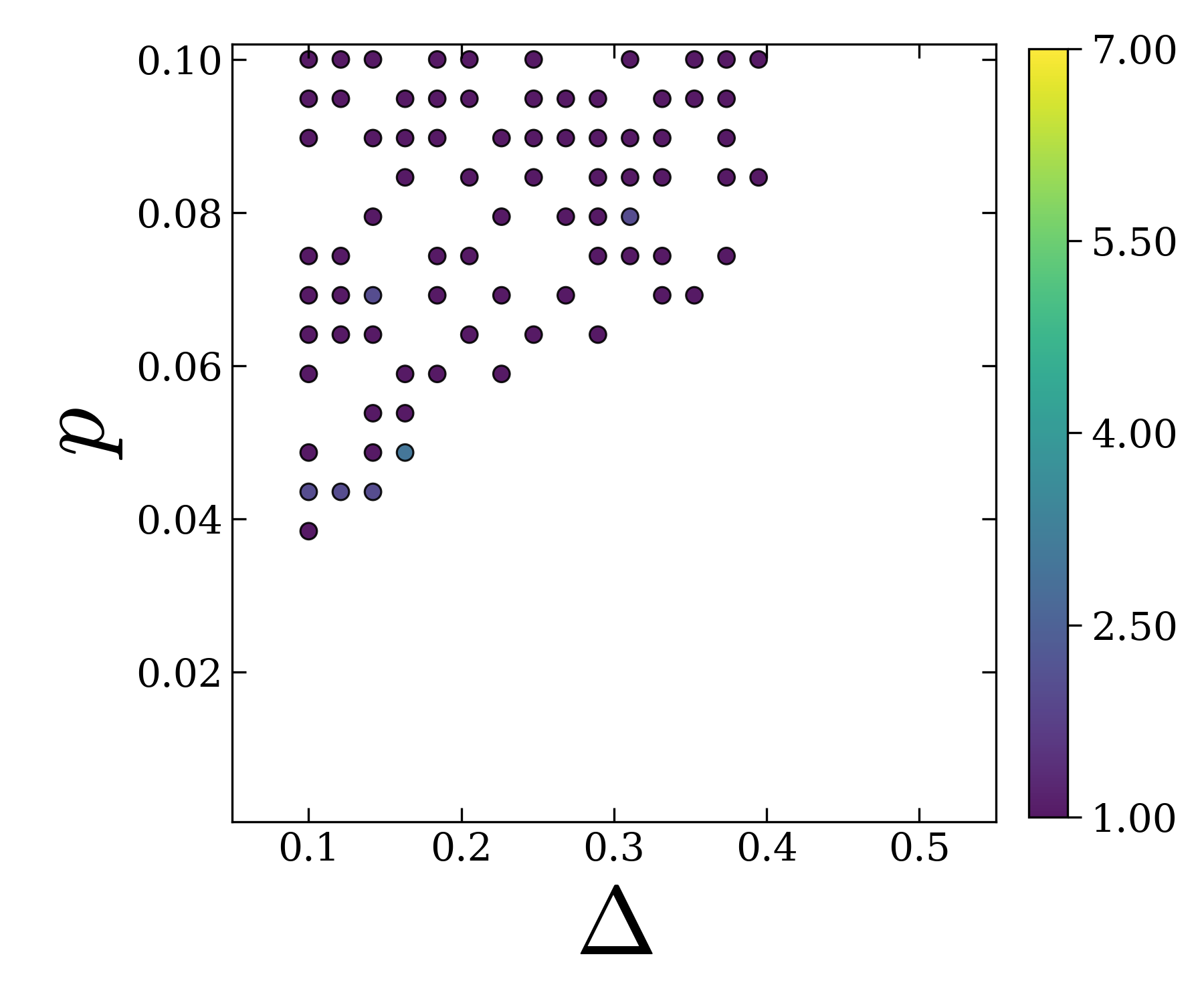}

\vspace{-0.15cm}

\makebox[0.32\textwidth]{\small (d)}
\hfill
\makebox[0.32\textwidth]{\small (e)}
\hfill
\makebox[0.32\textwidth]{\small (f)}

\caption{
Phase boundary and anomalous cluster visualisation for a network of
$N=100$ at $T=250$. The \textbf{top row} (a-c) corresponds to
$s=1.0$, while the \textbf{bottom row} (d-f) corresponds to
$s=0.5$. Panels (a,d) show the SVM phase boundary. Panels
(b,e) show the cluster formation in anomalous points, and
panels (c,f) show the cluster formation in non-anomalous
points. Increased connectivity raises the number of anomalous
points and enhances the extent of cluster formation. Colors represent the size of the clusters.
}

\label{fig:anom_100}
\end{figure*}

\section{Anomalous Cluster Persistence}

\subsection{Motivation and Hypothesis}

During SVM boundary validation, a subset of parameter points within the predicted
Binary Dominance Region exhibited unexpected coexistence (non-binary dominance).
We hypothesise, consistent with agent-based modelling theory \cite{castello2007anomalous},
that this anomaly arises from the formation of \emph{localised minority clusters}
--- connected subgraphs of minority speakers that are sufficiently isolated from
the majority to resist convergence.  The dynamic weighting rule (Eq.~\eqref{eq:weight})
reinforces intra-cluster edges over time, making these communities increasingly
resistant to external pressure\cite{patriarca2009influence}.

\subsection{Cluster Quantification at $T = 250$}

Fig.$4$ and Fig.$5$ illustrate the relationship between the SVM-predicted phase boundary and the formation of minority-language clusters for network sizes $N=50$ and $N=100$, respectively, after $T=250$ iterations. In both figures, the first row corresponds to the symmetric prestige case ($s=1.0$), while the second row shows the results for $s=0.5$, allowing the influence of language prestige on cluster formation to be examined.

A common feature observed in both network sizes is that the anomalous non-binary dominant states are concentrated near the SVM phase boundary. These anomalous points consistently coincide with the presence of compact and well-connected minority-language clusters, indicating that adaptive edge reinforcement enables locally connected minority communities to resist complete extinction even within regions where the SVM predicts majority-language dominance. In contrast, non-anomalous parameter points exhibit either fragmented or very small minority-language clusters, allowing the majority language to propagate throughout the network and ultimately achieve global dominance. This comparison demonstrates that the anomalous coexistence originates from dynamically reinforced minority communities rather than from inaccuracies in the SVM classification.

The comparison between the two prestige values further shows that the underlying clustering mechanism remains qualitatively unchanged. Although the spatial distribution and density of anomalous points vary between $s=1.0$ and $s=0.5$, minority-language persistence is consistently associated with the survival of locally reinforced clusters. Thus, varying the prestige modifies the extent of the anomalous coexistence region without changing the physical mechanism responsible for its formation.

A comparison of Fig.$4$ and Fig.$5$ highlights the influence of system size on cluster persistence. While both network sizes exhibit anomalous coexistence near the phase boundary, the larger network ($N=100$) contains a greater number of anomalous configurations together with more extensive cluster formation. The increased number of agents provides additional opportunities for minority-language speakers to establish locally connected communities before the majority language spreads throughout the network. Consequently, dynamically reinforced clusters survive over a broader region of the parameter space than in the smaller system.

Overall, it provide direct microscopic evidence for the origin of the anomalous coexistence region identified by the SVM analysis. Rather than representing classification errors, the anomalous states arise from the formation of compact minority-language clusters that are stabilised through adaptive edge reinforcement. These results establish a clear connection between the macroscopic phase boundary and the underlying cluster dynamics responsible for minority-language persistence.

\begin{figure*}[t]
\centering

\includegraphics[width=0.32\textwidth]{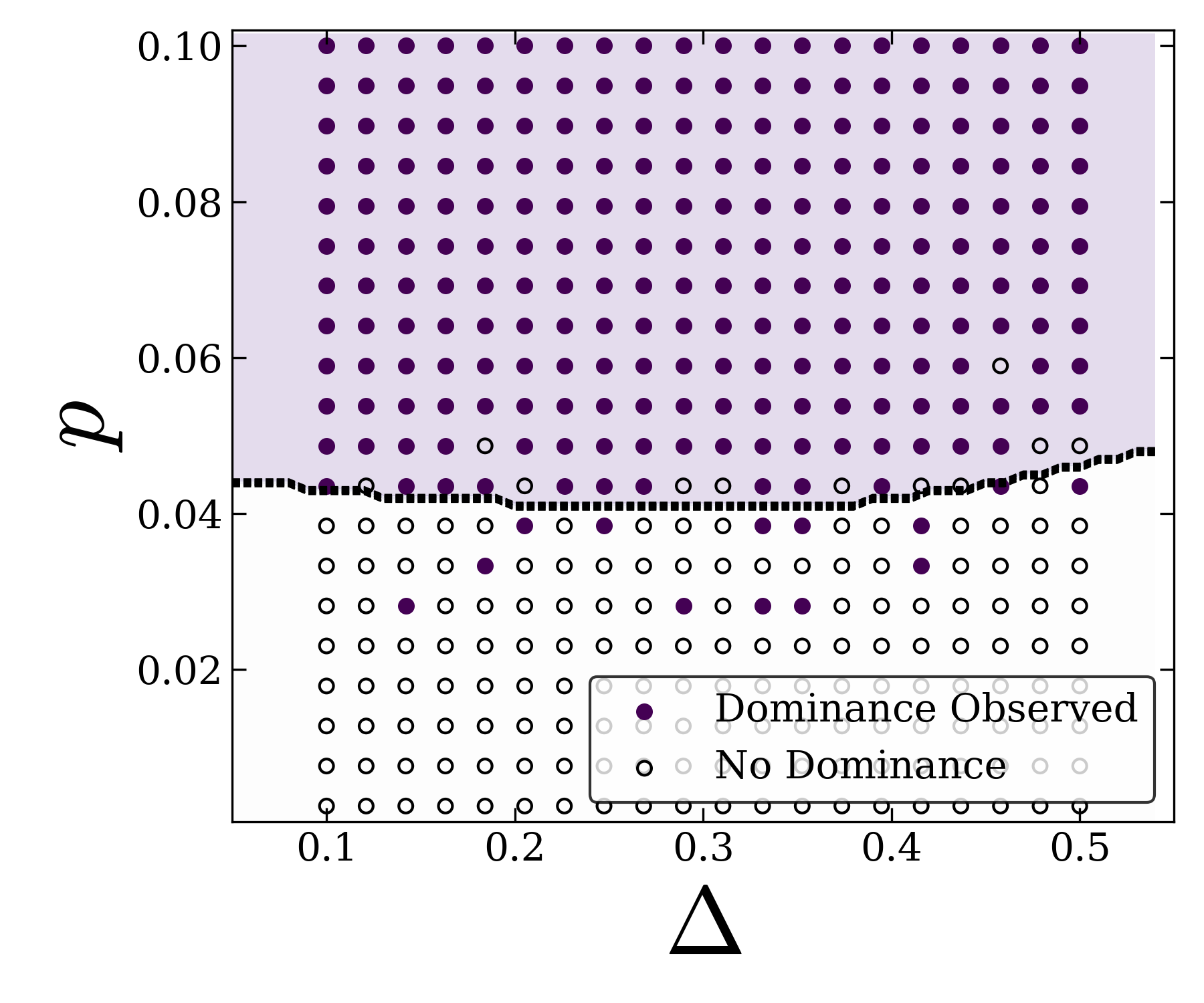}
\hfill
\includegraphics[width=0.32\textwidth]{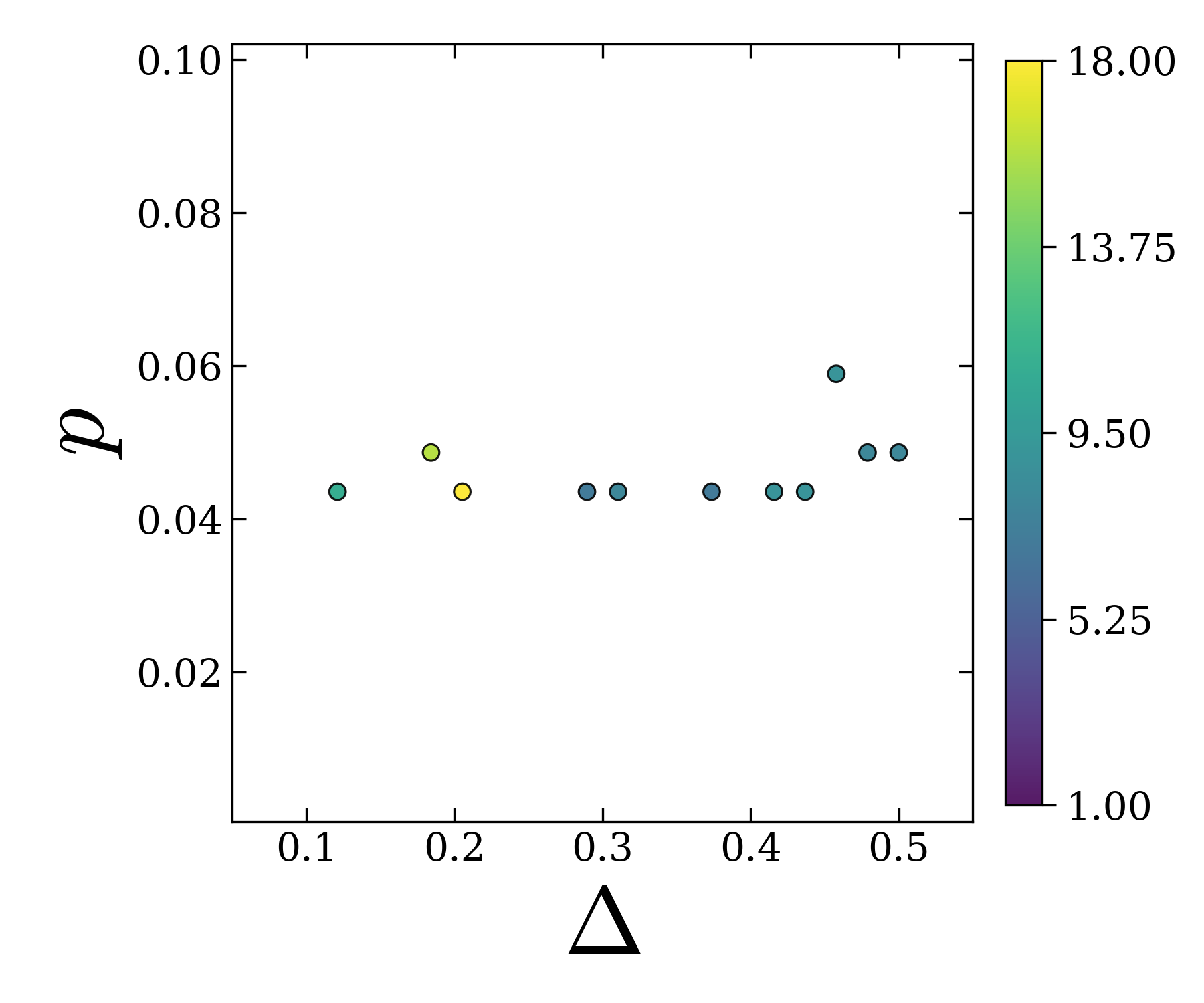}
\hfill
\includegraphics[width=0.32\textwidth]{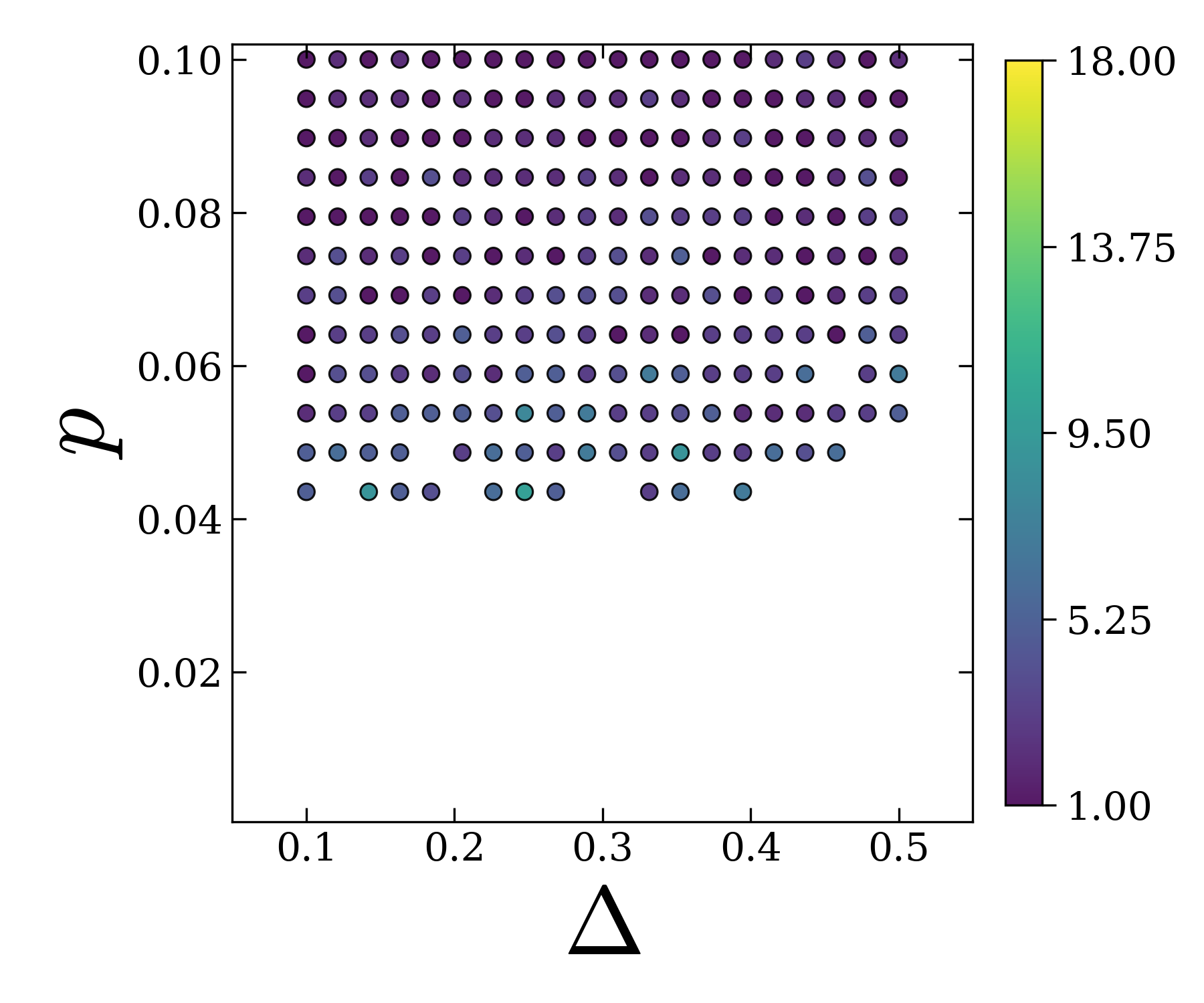}

\vspace{-0.15cm}

\makebox[0.32\textwidth]{\small (a)}
\hfill
\makebox[0.32\textwidth]{\small (b)}
\hfill
\makebox[0.32\textwidth]{\small (c)}
\hfill

\vspace{0.15cm}

\includegraphics[width=0.32\textwidth]{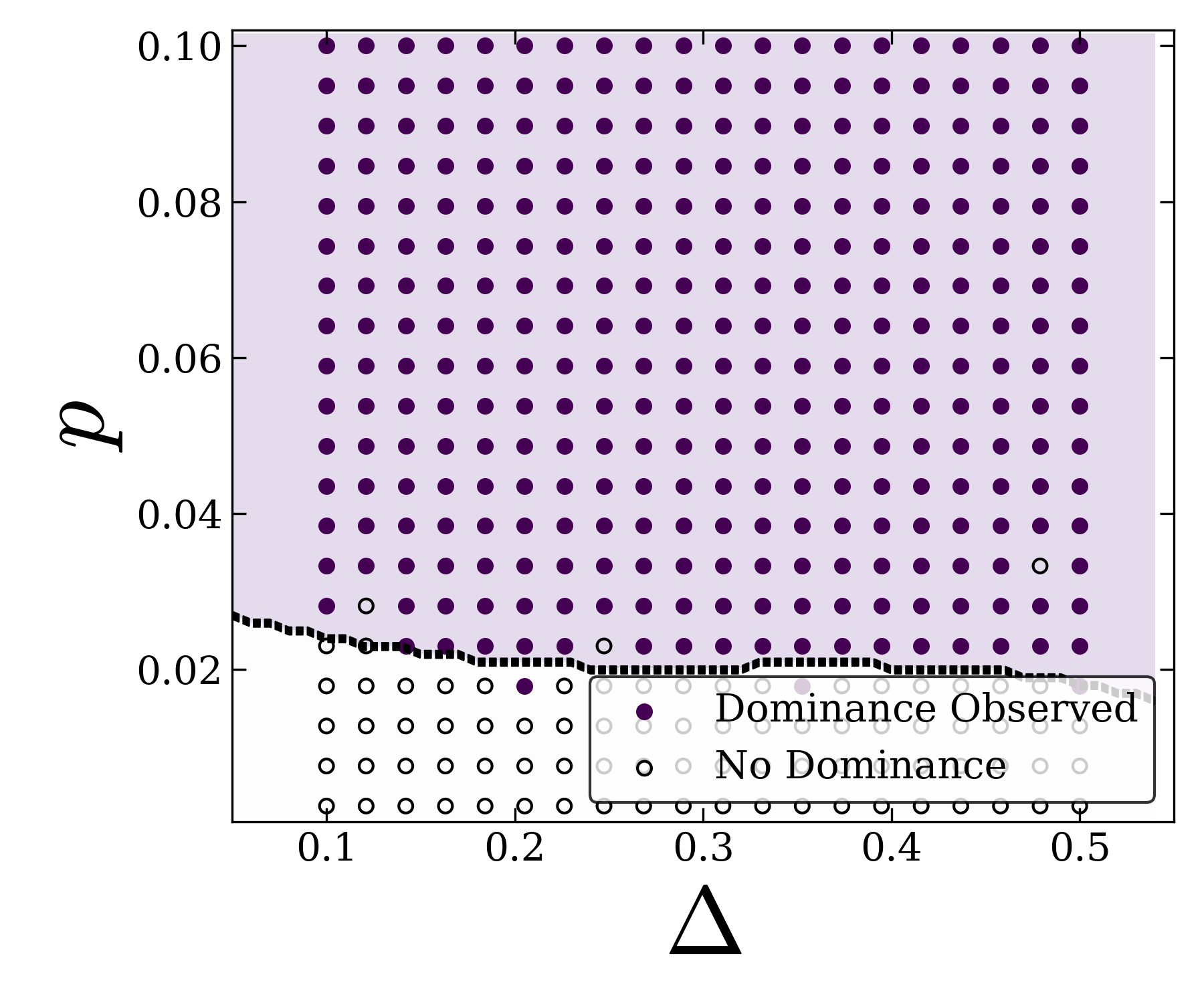}
\hfill
\includegraphics[width=0.32\textwidth]{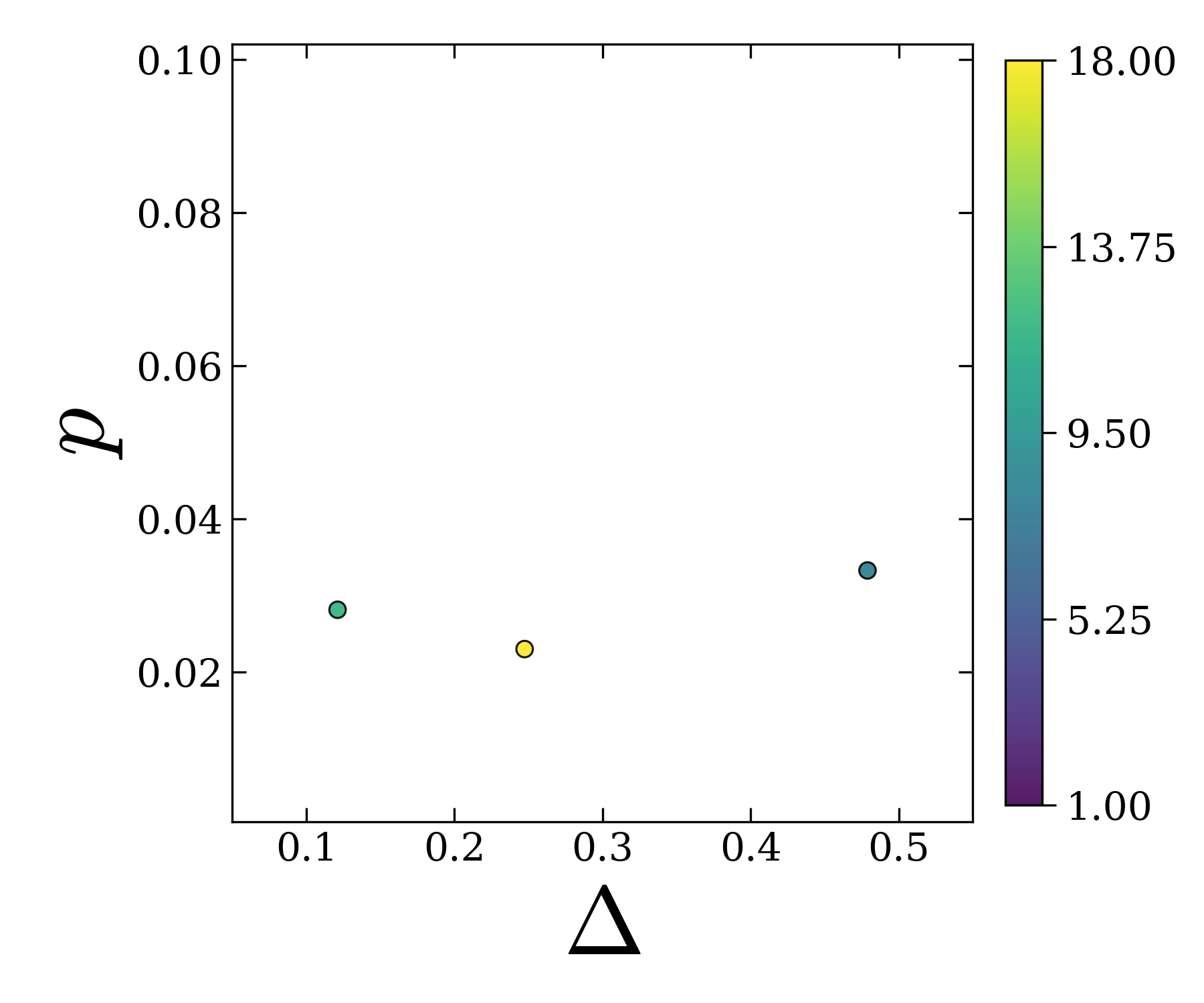}
\hfill
\includegraphics[width=0.32\textwidth]{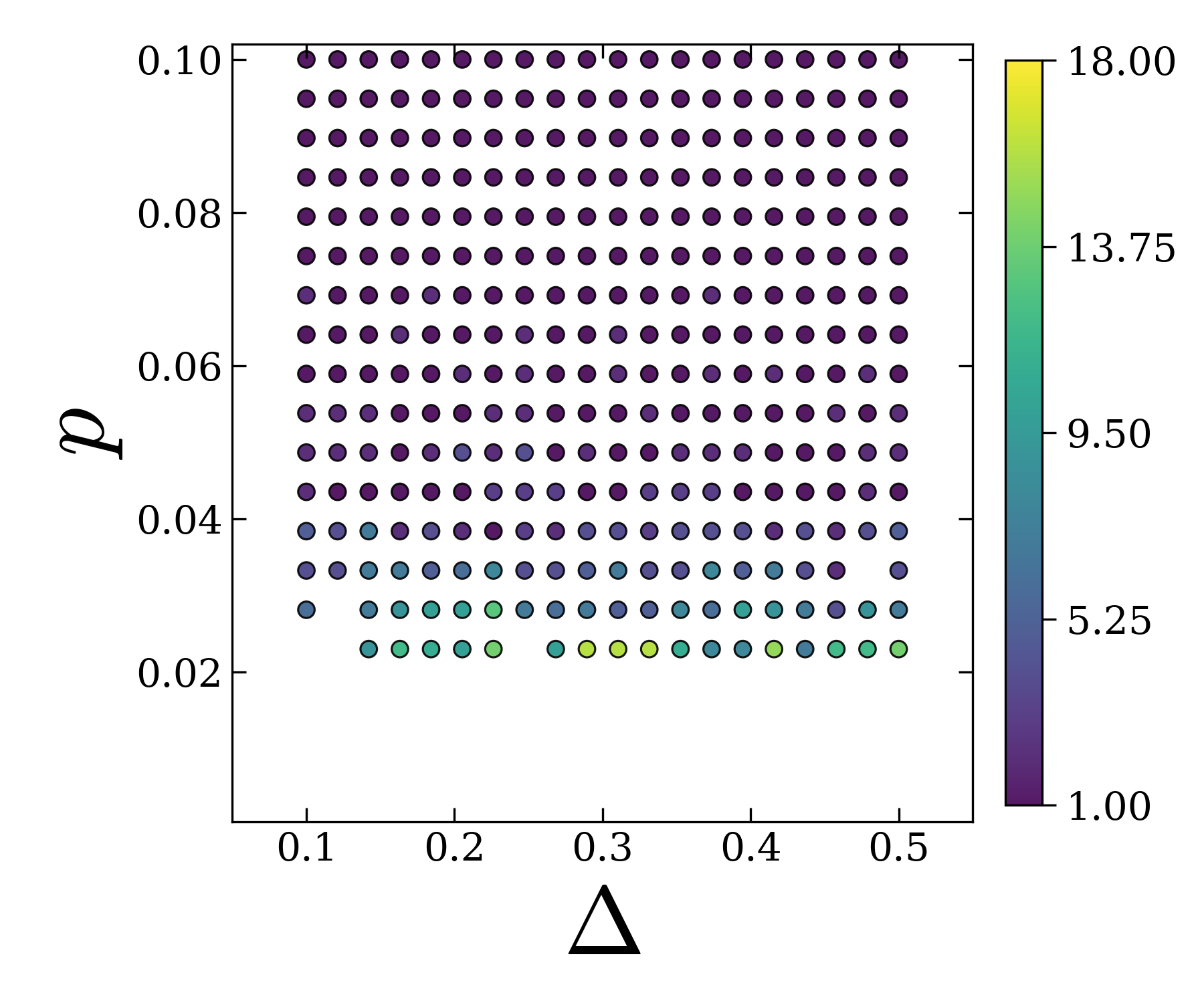}

\vspace{-0.15cm}

\makebox[0.32\textwidth]{\small (d)}
\hfill
\makebox[0.32\textwidth]{\small (e)}
\hfill
\makebox[0.32\textwidth]{\small (f)}
\hfill

\caption{
Phase boundary and anomalous cluster analysis at $T=750$ for two network
sizes. The \textbf{top row} (a-c) corresponds to $N=50$, while the
\textbf{bottom row} (d-f) corresponds to $N=100$. (a,d) shows SVM phase boundary. (b,e) show the cluster formation in anamalous points . Panel(c,f) shows the cluster formation in non anamalous points. 
}

\label{fig:anom_750}
\end{figure*}

\begin{figure*}[t]
\centering

\includegraphics[width=0.32\textwidth]{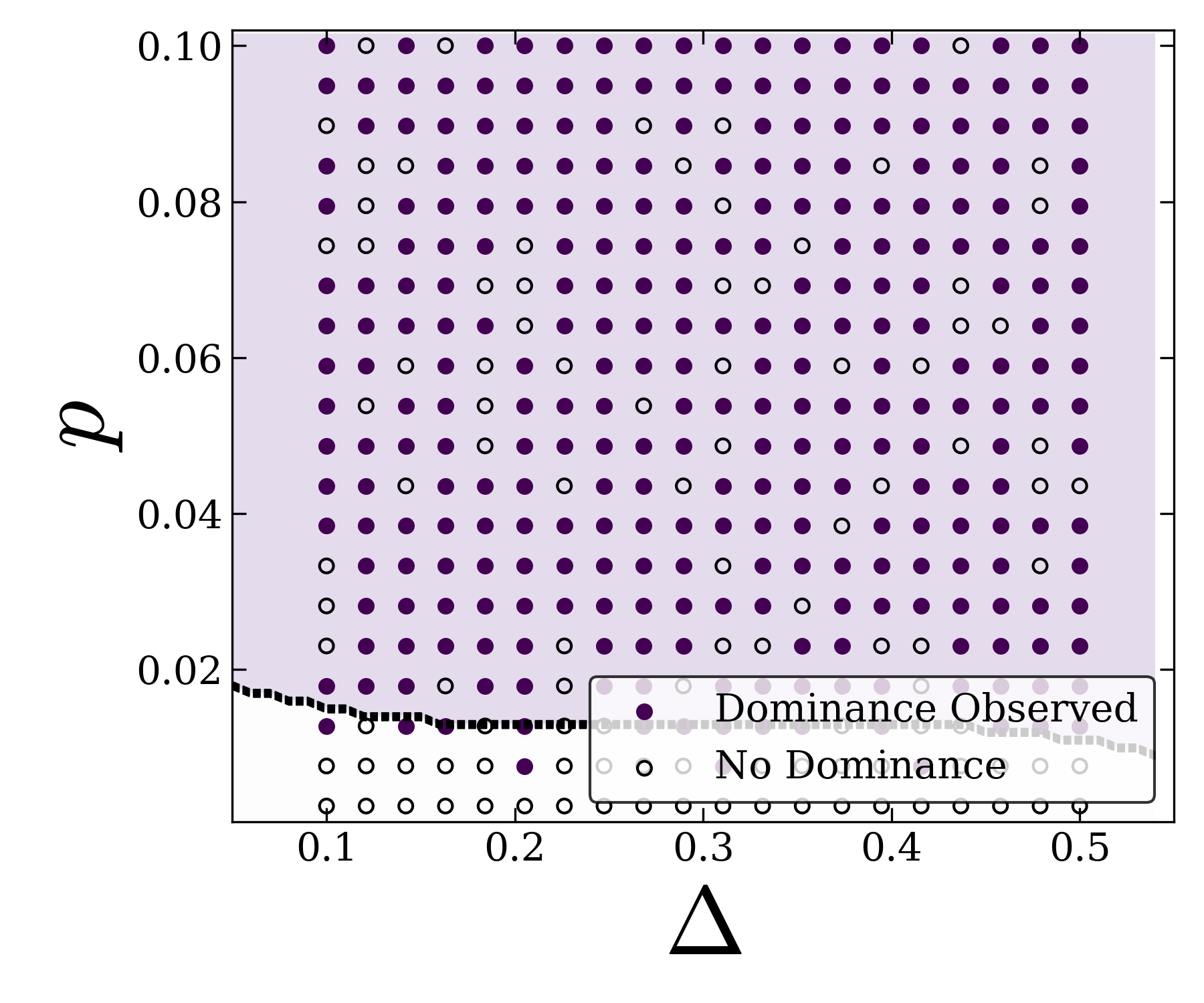}
\hfill
\includegraphics[width=0.32\textwidth]{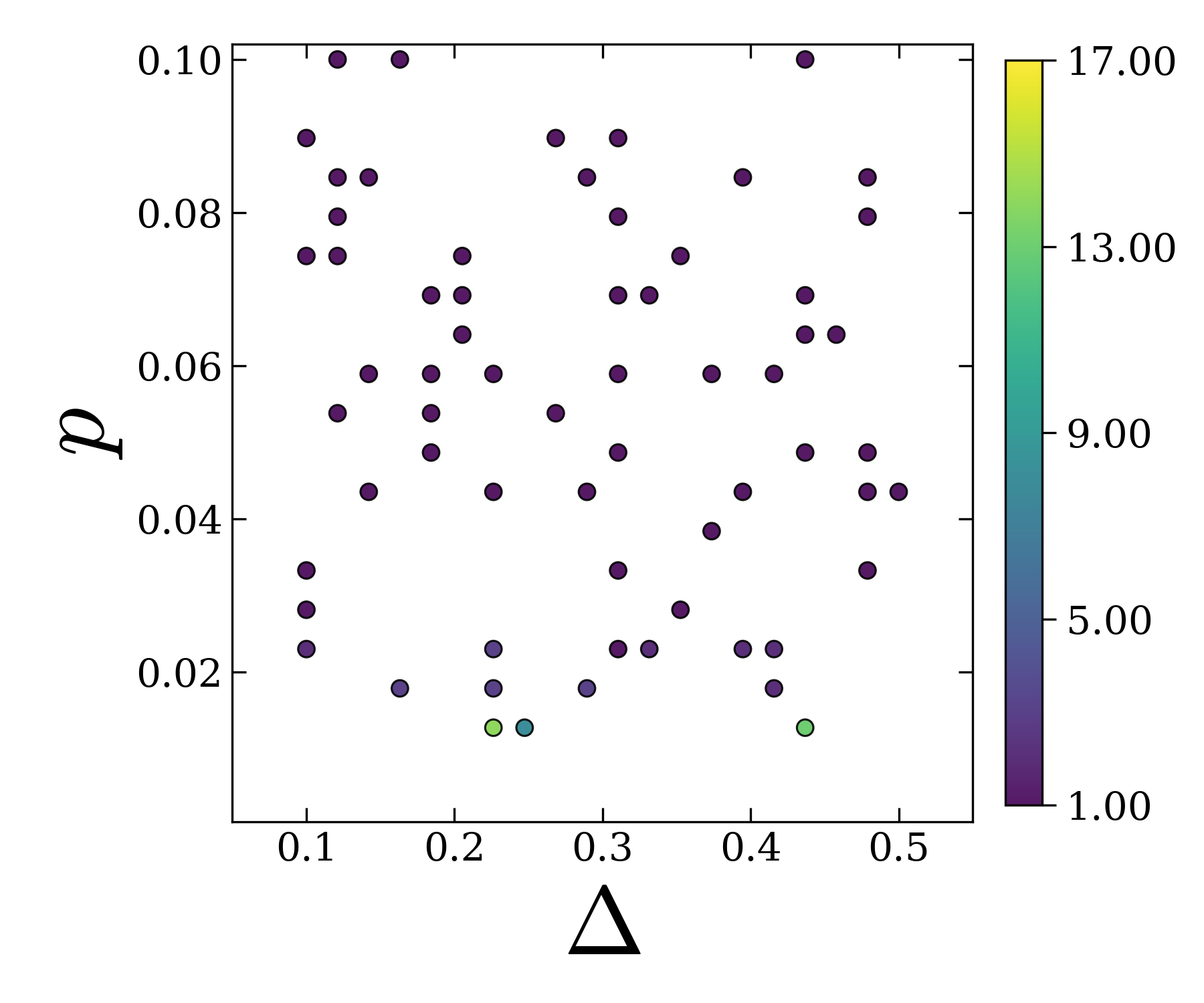}
\hfill
\includegraphics[width=0.32\textwidth]{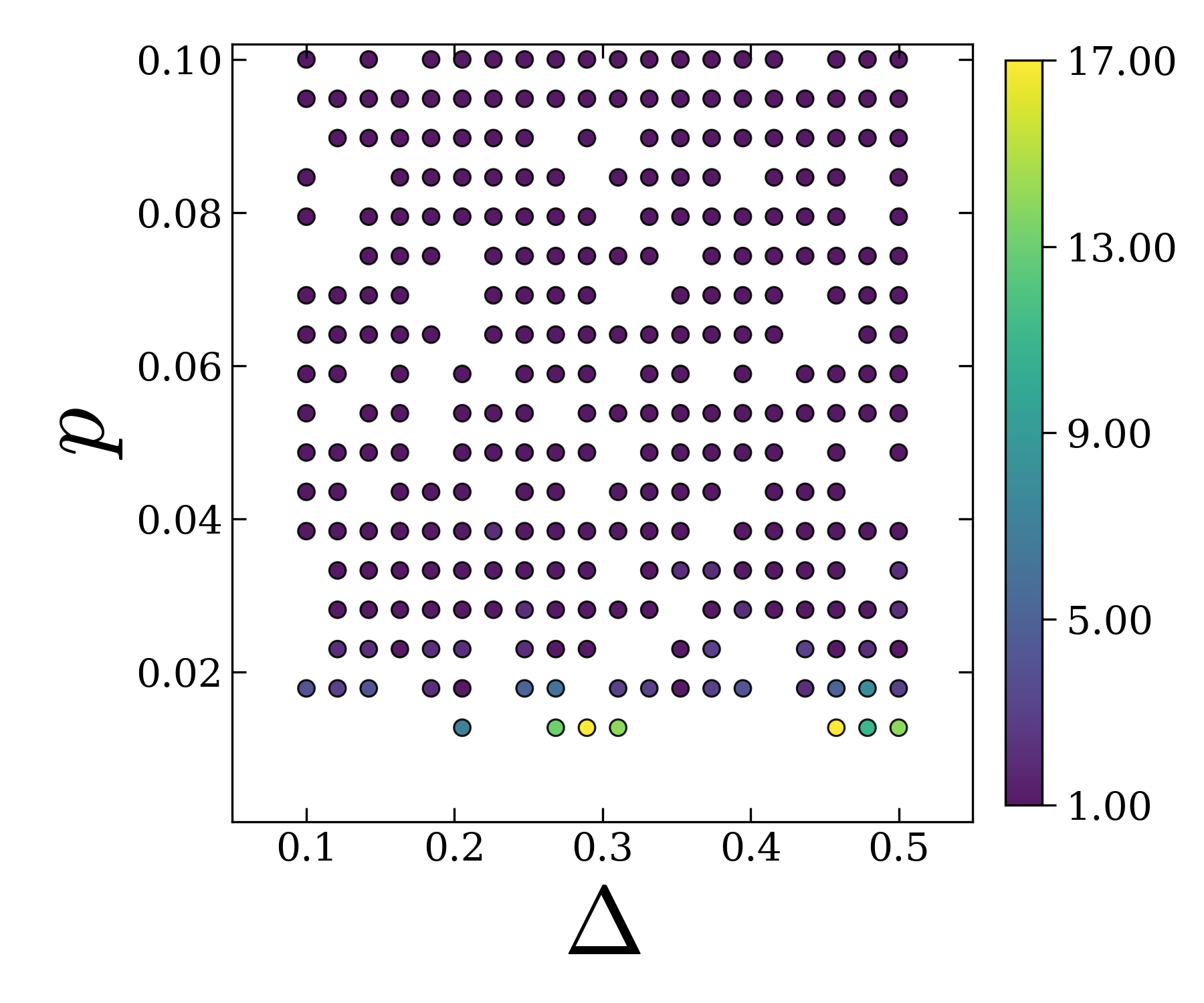}

\vspace{-0.15cm}

\makebox[0.32\textwidth]{\small (a)}
\hfill
\makebox[0.32\textwidth]{\small (b)}
\hfill
\makebox[0.32\textwidth]{\small (c)}
\hfill

\vspace{0.15cm}

\includegraphics[width=0.32\textwidth]{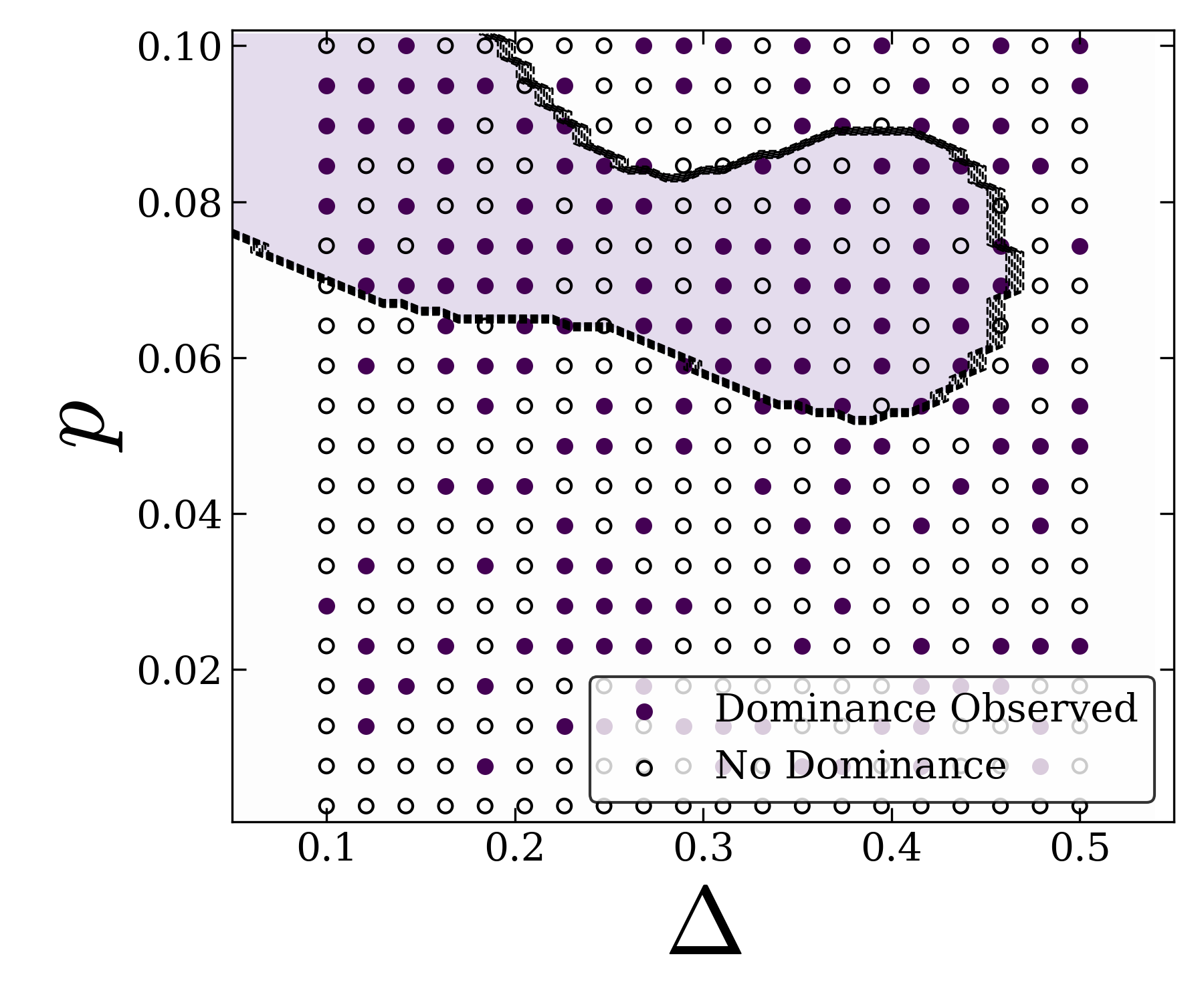}
\hfill
\includegraphics[width=0.32\textwidth]{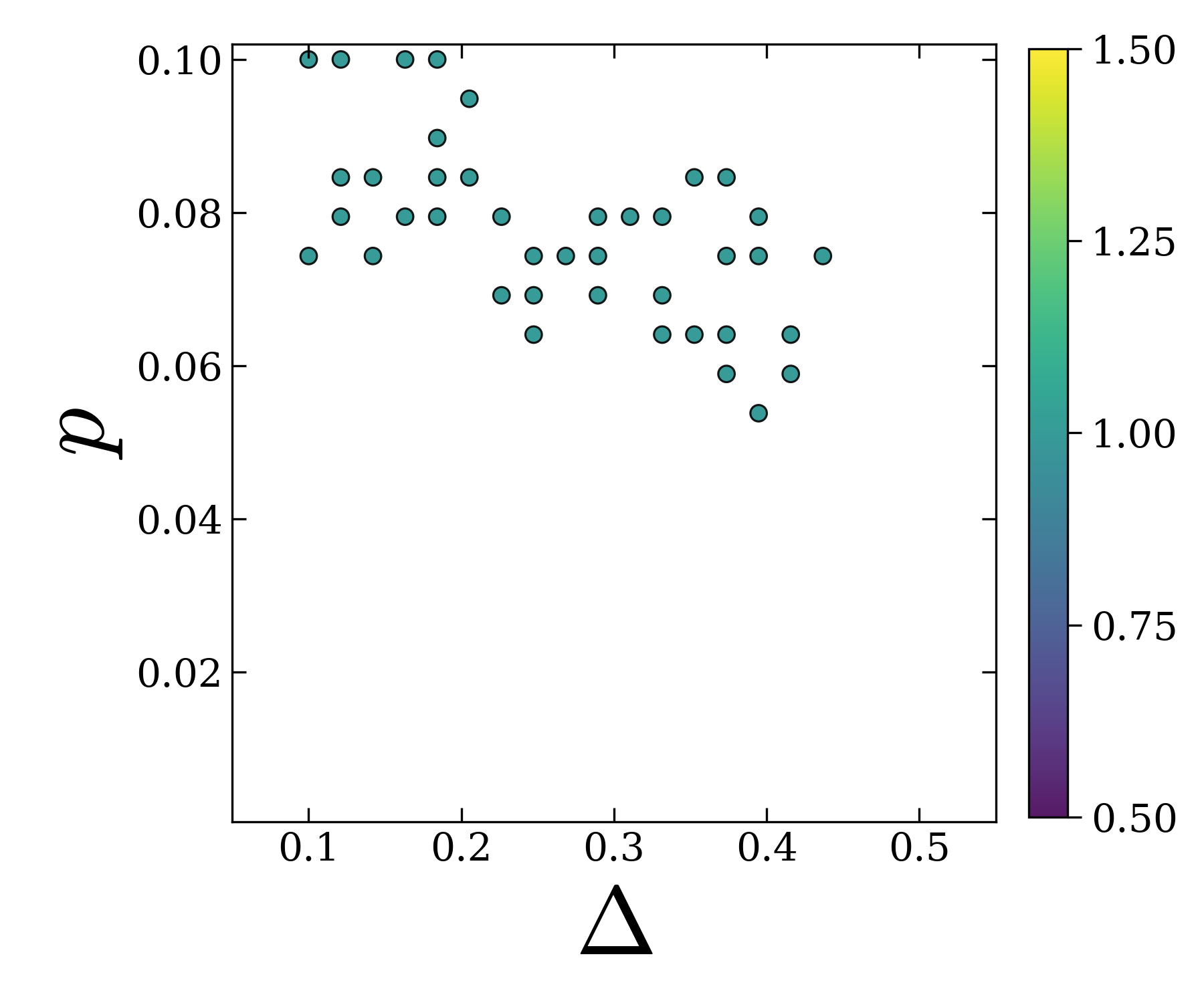}
\hfill
\includegraphics[width=0.32\textwidth]{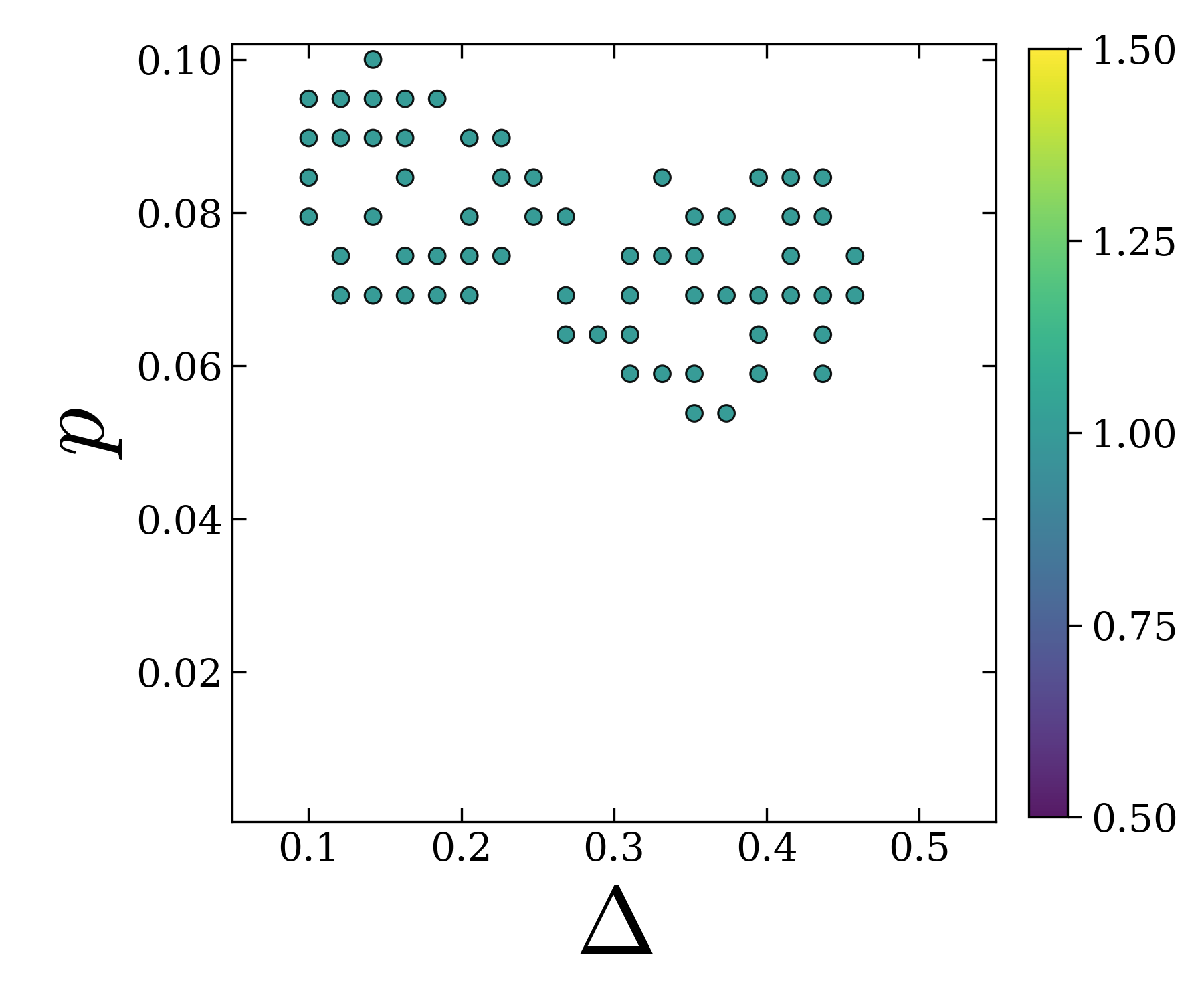}

\vspace{-0.15cm}

\makebox[0.32\textwidth]{\small (d)}
\hfill
\makebox[0.32\textwidth]{\small (e)}
\hfill
\makebox[0.32\textwidth]{\small (f)}
\hfill                 
\caption{
Phase boundary and anomalous cluster analysis at $T=750$ for network
sizes $N=250$ and $N=500$. The \textbf{top row} (a-c) corresponds to
$N=250$, while the \textbf{bottom row}(d-f) corresponds to $N=500$.
(a,d) shows the SVM phase boundary, where open white circles denote non-binary dominant (anamalous) points within the predicted dominance region. panel (b,e) show the cluster formation in anamalous points, (c,f) show the cluster formation in non-anamalous points.
}

\label{fig:anom_250_500_750}
\end{figure*}

\subsection{Cluster Quantification at $T = 750$}

To investigate the temporal evolution of anomalous clustering, we repeat the above
analysis with an extended simulation horizon of $T = 750$ iterations.  The following
phase-space plots were generated under identical parameter grids and network
sizes, but with three times as many evolution steps.

To investigate the long-time stability of anomalous coexistence, the simulations were extended to $T=750$ iterations while maintaining the same parameter space and network configurations used for the $T=250$ analysis. Fig.$6$ and Fig.$7$ present the corresponding phase-boundary and cluster analyses for network sizes ranging from $N=50$ to $N=500$. These results enable a direct comparison between intermediate-time and long-time dynamics, providing insight into the persistence and evolution of minority-language clusters.

Across all network sizes, the anomalous non-binary dominant states remain concentrated near the SVM phase boundary, indicating that the underlying mechanism responsible for anomalous coexistence is preserved during long-time evolution. Minority-language persistence continues to arise from dynamically reinforced local clusters that temporarily withstand the surrounding majority-language influence. However, the number of anomalous configurations decreases significantly compared with the results obtained at $T=250$, demonstrating that many coexistence states observed at shorter times are metastable rather than true steady-state solutions. As the dynamics proceeds, weakly connected minority communities gradually dissolve, allowing the majority language to dominate a larger fraction of the parameter space.

The effect of network size becomes more pronounced during the extended evolution. Although the qualitative phase behaviour remains consistent across all values of $N$, larger systems exhibit fewer persistent anomalous states together with a lower mean cluster count than smaller networks. The increased number of communication pathways in larger networks facilitates the spread of the majority language, making long-term topological isolation of minority-language communities increasingly difficult. Consequently, the surviving minority clusters become more compact and localized, particularly for $N=250$ and $N=500$, where most anomalous configurations are characterised by only one or a few isolated minority clusters.

A comparison between the $T=250$ and $T=750$ results therefore reveals a systematic temporal evolution of the phase behaviour. Increasing the simulation time sharpens the phase boundary by eliminating transient coexistence states, while increasing the network size further suppresses minority-language persistence through enhanced global connectivity. Despite these quantitative changes, the microscopic mechanism responsible for anomalous coexistence remains unchanged: minority-language survival is consistently associated with dynamically reinforced clusters located near the phase boundary. The extended simulations therefore demonstrate that adaptive edge reinforcement can substantially delay language extinction but cannot indefinitely prevent majority-language dominance unless sufficiently robust and well-connected minority communities are established.

Two robust trends emerge when comparing the $T = 250$ and $T = 750$ results:

{\textbf{a. Clustering decreases with simulation time:}
At $T = 750$ the anomalous points (non-dominant outcomes inside the predicted
dominance region) are fewer and less scattered than at $T = 250$ for every network
size tested. 
The dynamic reinforcement eventually builds sufficient intra-cluster cohesion for
minority clusters to either consolidate into stable enclaves or, more commonly,
succumb to majority pressure once their periphery is eroded.  In other words,
extended simulation reveals that many of the anomalous points at $T = 250$ were
\emph{transient} --- the system had not yet reached its absorbing state within
250 steps.

{\textbf{b. Clustering decreases with network size}:}
Across both time horizons, the density of anomalous points and the average
cluster count in non-dominant trials decrease as $N$ increases from 50 to 500.
Larger networks provide more alternative majority-language paths around any
minority enclave, making topological isolation harder to sustain.  This is in
line with the thermodynamic-limit argument: in the limit $N \to \infty$ the
phase boundary sharpens and the measure of the anomalous transition zone shrinks
to zero \cite{castello2007anomalous,vazquez2010agent}.

\begin{table*}[t]
\centering
\caption{Summary of scaling behaviour across network sizes.}
\label{tab:scaling}

\begin{tabular}{ccccc}
\toprule
$N$ & Average cluster size & Phase boundary width &
Anomaly density & Iterations ($T$) \\
\midrule
50   & High (10 - 18)   & Sharp  & Low (4.64\%)    & 750 \\
100  & High (9 - 13.5) & Medium & Low (1.25\%) & 750 \\
250  & Medium (5 - 17) & Medium & Medium (18.82\%) & 750 \\
500  & Low (1 - 1.5)    & Broad  & High (39.21 \%)   & 750 \\
\bottomrule
\end{tabular}

\end{table*}

\subsection{Scaling Analysis and Key Insights}

Three scaling laws emerge from the above study:

\textbf{a. Temporal evolution of phase stability.}
Over time, the dynamic edge reinforcement causes minority clusters to become
internally more cohesive. The phase boundary sharpens as the system approaches
steady state, and the misclassification rate of the SVM decreases as $T$ grows.

\textbf{b. Impact of system scale ($N$).}
As $N$ increases, the average minority cluster size decreases relative to $N$.
Larger systems exhibit greater stochastic noise at the majority--minority interface,
causing the effective phase boundary to broaden compared to smaller, more
tightly-coupled networks.

\textbf{c. Boundary dynamics.}
The density of anomalous (non-binary dominant) points is highest in the $(p, \Delta)$
region proximal to the SVM boundary. This indicates that the dominance-to-coexistence
transition is governed by the fragmentation of the population into a larger number
of smaller, isolated minority clusters, consistent with the community-structure
findings of \cite{castello2007anomalous}.

Table 2 summarises the qualitative scaling behaviour observed across
different network sizes at $T=750$. As the system size increases, the average cluster size --
the number of nodes in a connected minority-language subgraph surviving at an anomalous
parameter point -- decreases, falling from $10$--$18$ nodes at $N=50$ to $1$--$1.5$ nodes at
$N=500$, indicating that larger networks favour the fragmentation of minority-language
communities into smaller isolated groups. At the same time, the phase boundary width, which
describes how cleanly the SVM boundary separates the dominance and coexistence regimes,
becomes progressively broader, reflecting the increasing influence of stochastic fluctuations in
larger systems. This broadening is accompanied by an increase in the anomaly density -- the
percentage of $(p,\Delta)$ grid points classified as Dominance by the SVM that instead exhibit
non-binary (coexistence) behaviour in simulation -- rising from $4.64\%$ at $N=50$ to $39.21\%$
at $N=500$, suggesting that finite-size effects become more pronounced as the number of agents
increases. Despite these quantitative changes, the underlying mechanism responsible for
anomalous coexistence remains unchanged. Minority-language persistence continues to arise
from the formation of compact, dynamically reinforced clusters that survive within the
SVM-predicted dominance region. These observations demonstrate that network size primarily
affects the sharpness of the phase transition and the spatial distribution of anomalous states,
while the fundamental role of adaptive edge reinforcement in sustaining minority-language
communities is preserved across all system sizes.

\begin{figure*}[t]
  \centering
  \includegraphics[width=0.32\textwidth]{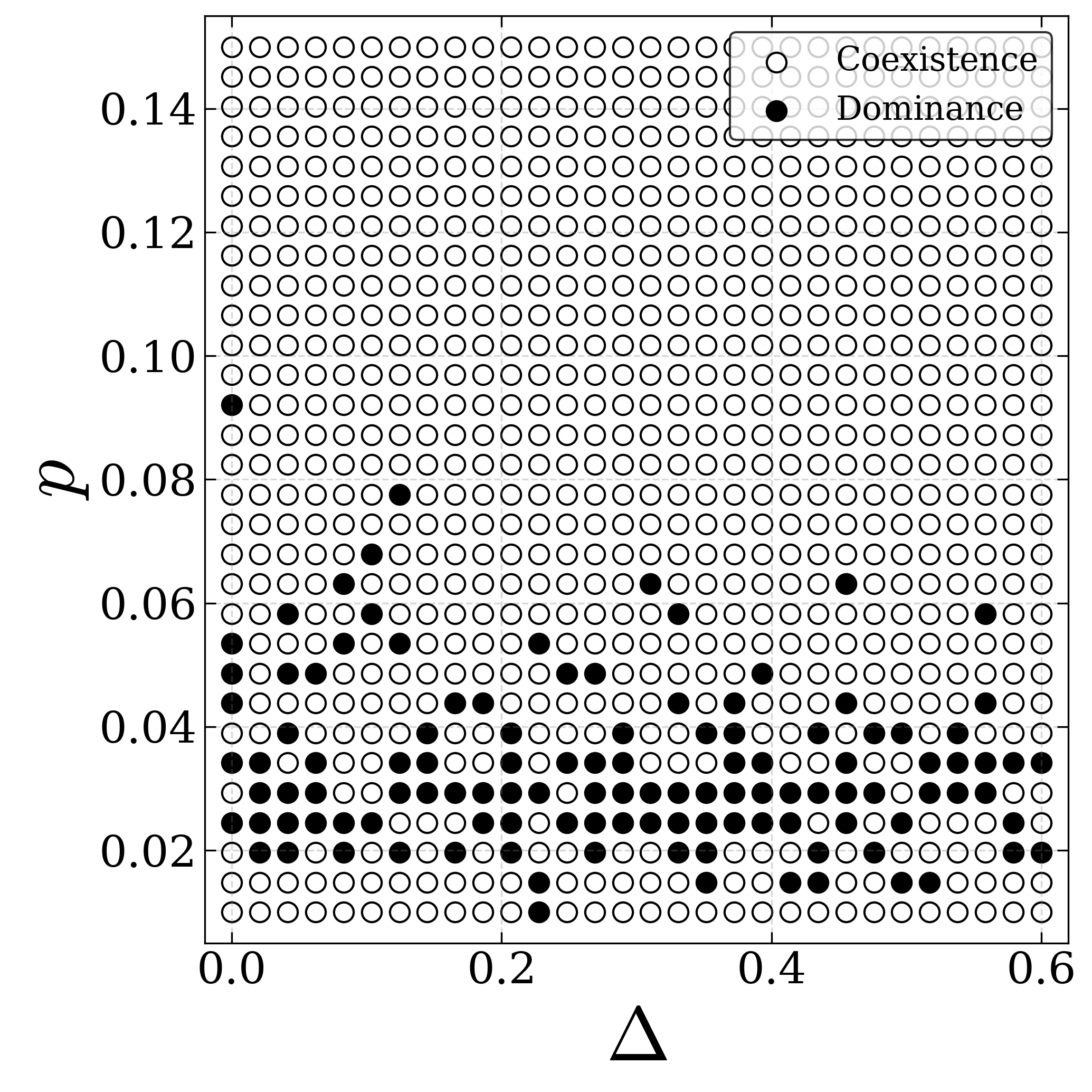}
  \hfill
  \includegraphics[width=0.32\textwidth]{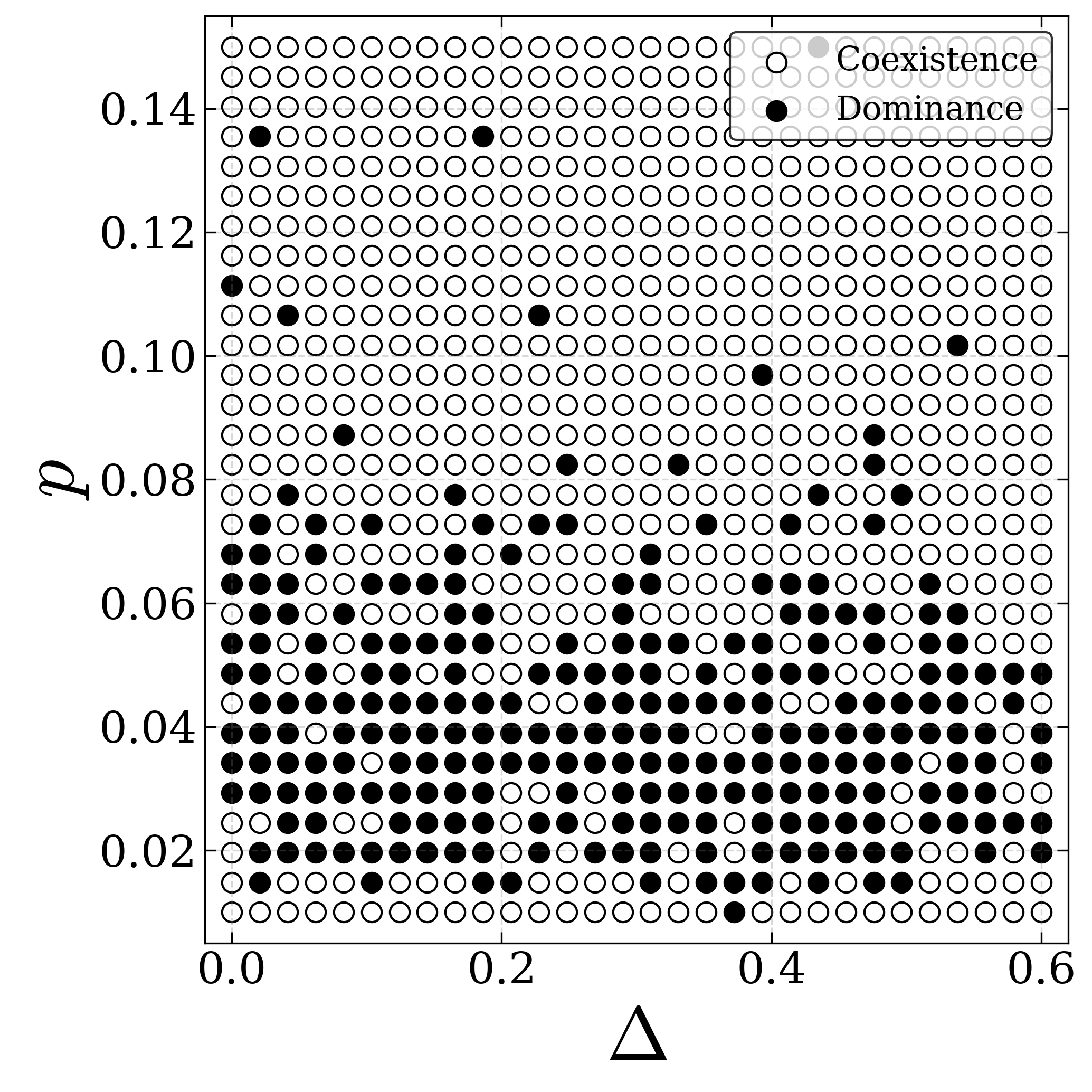}
  \hfill
  \includegraphics[width=0.32\textwidth]{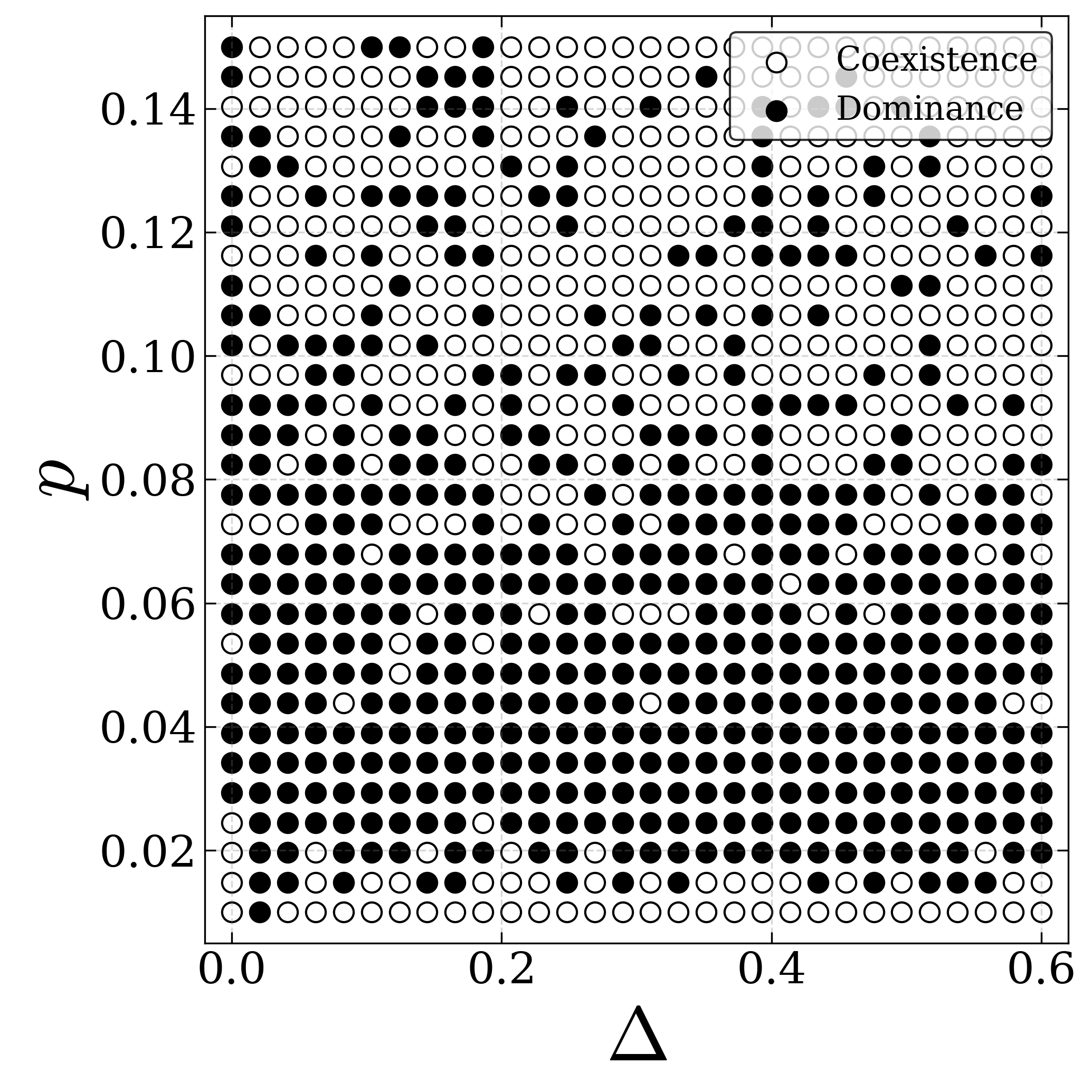}  
  
   \vspace{-0.15cm}

   \makebox[0.32\textwidth]{\small (a)}
    \hfill
    \makebox[0.32\textwidth]{\small (b)}
    \hfill
    \makebox[0.32\textwidth]{\small (c)}
    
     \vspace{0.15cm}
  \includegraphics[width=0.32\textwidth]{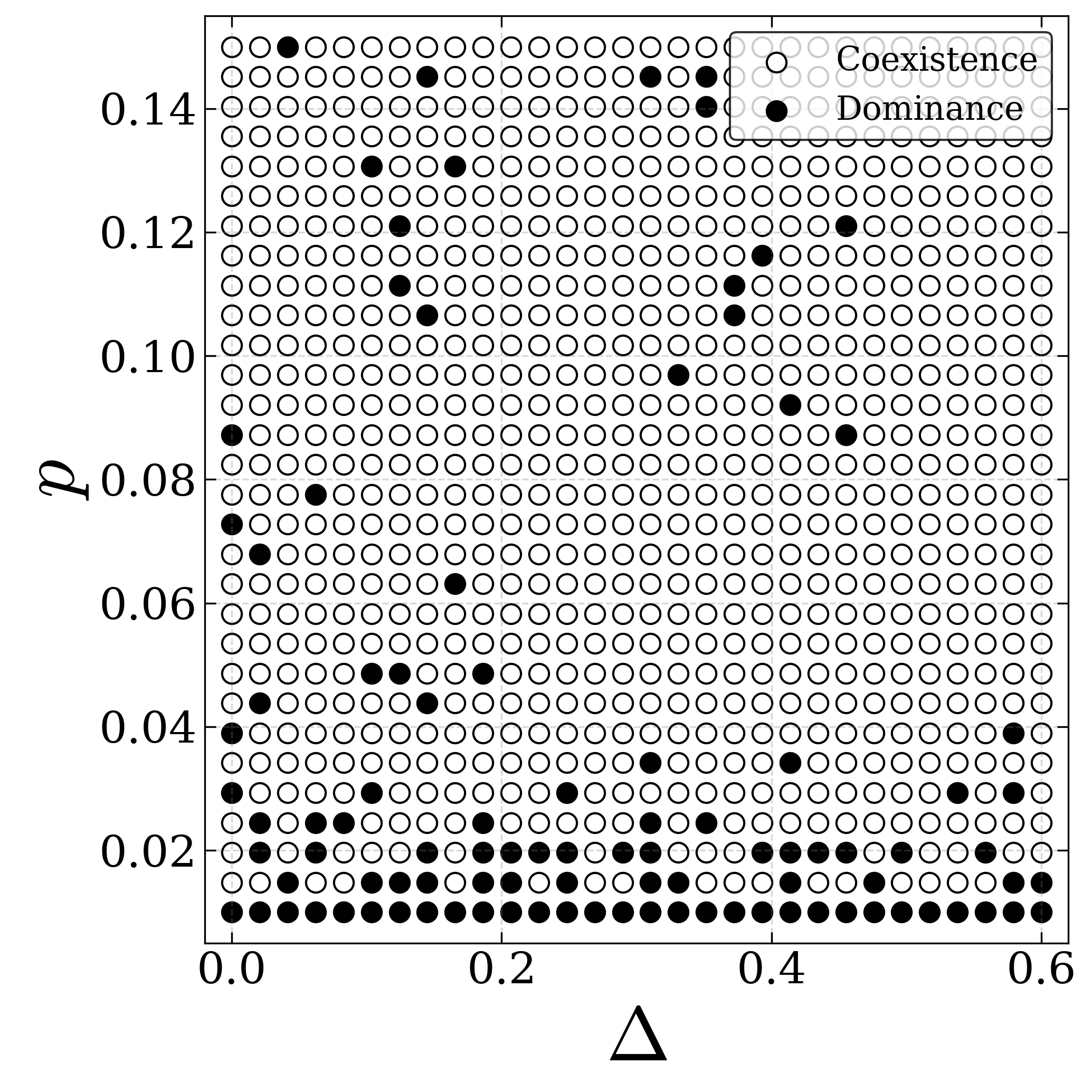}
  \hfill
  \includegraphics[width=0.32\textwidth]{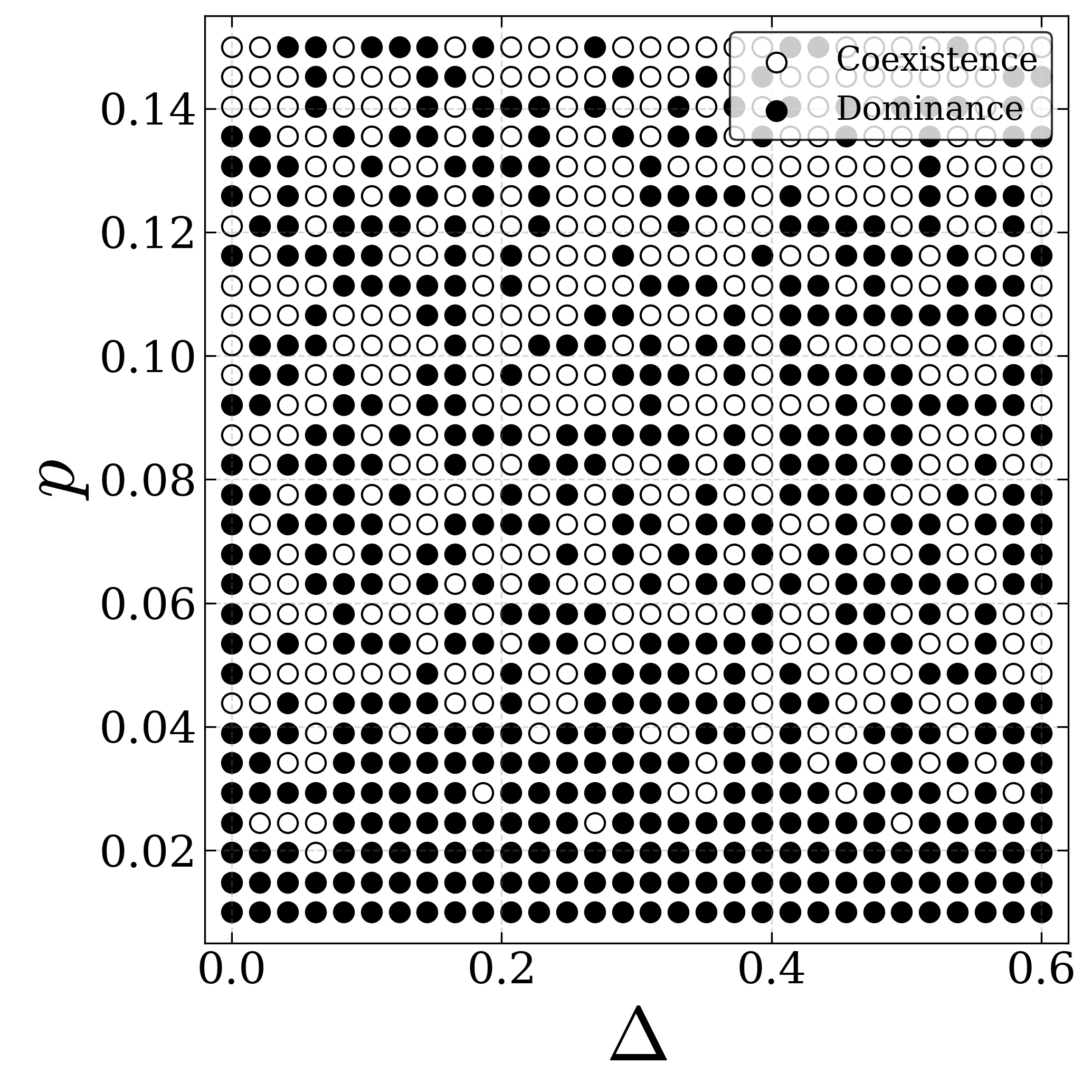}
  \hfill
  \includegraphics[width=0.32\textwidth]{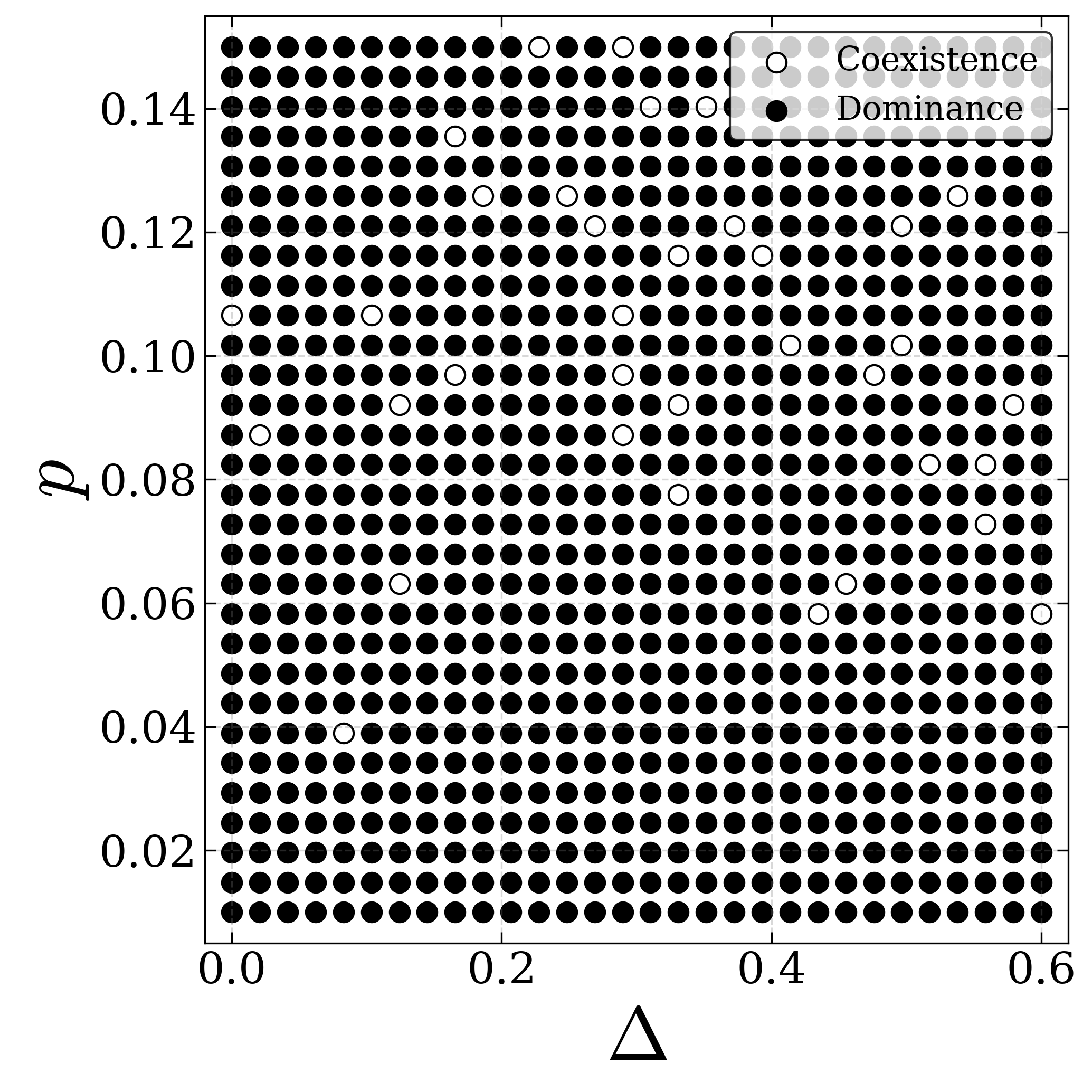}
    \vspace{-0.15cm}
  
   \makebox[0.32\textwidth]{\small (d)}
   \hfill
   \makebox[0.32\textwidth]{\small (e)}
   \hfill
\makebox[0.32\textwidth]{\small (f)}

  \caption{Phase diagrams for the bilingual model extension.
           \emph{Top row} (100 nodes): (a)~$s = 0.63$ (Coexistence);
           (b)~$s = 0.66$ (Critical); (c)~$s = 0.67$ (Dominance).
           \emph{Bottom row} (1000 nodes): (d)~$s = 0.73$ (Coexistence);
           (e)~$s = 0.75$ (Critical); (f)~$s = 0.77$ (Dominance).
           The transition from coexistence through the critical point to
           monolingual dominance as prestige $s$ increases is visible in both rows.}
  \label{fig:bilingual}
\end{figure*}

\section{Discussion}

The results presented here extend the Abrams--Strogatz framework in two complementary
directions. First, the dynamic edge-weighting mechanism provides a principled way to
encode the social bonding component of language loyalty: minority speakers who interact
frequently become increasingly resistant to majority pressure. This is qualitatively
consistent with \cite{milroy1987language}'s finding that dense, multiplex social networks
confer greater language vitality on minority varieties.

Second, the SVM-based phase boundary identification provides a data-driven complement
to the analytic fixed-point analysis of \cite{vazquez2010agent}. While analytic methods
are exact but often limited to simple network topologies, the SVM approach scales
naturally to complex parameter landscapes generated by simulation, and provides a
classification framework that can incorporate additional features (e.g., degree
heterogeneity, clustering coefficient) in future work.

The anomalous cluster persistence results suggest a new mechanism for minority language
survival that operates even in regimes where mean-field theory predicts extinction:
topological trapping of minority speakers in loosely connected subgraphs. This is
analogous to the ``topological traps'' described by \cite{castello2007anomalous} for
community-structured networks and suggests that the $\Delta$-reinforcement mechanism
may amplify these trapping effects.

A limitation of the current study is that the simulations use a fixed $T = 250$ for
all $N$, which means larger networks may not have reached steady state.  Future work
will use adaptive stopping criteria (e.g., when the fraction of switching agents falls
below a threshold) to ensure comparability across scales.

\section{Extension to Bilingual Models}
The binary $\{A, B\}$ framework studied above omits an important real-world category:
bilingual speakers who actively use both languages.  Following \cite{castello2006ordering},
we outline an extension to a tripartite state space $\{A, AB, B\}$, where $AB$ is
the bilingual state.

\subsection{Bilingual Transition Probabilities}

The extended transition probabilities are:
\begin{align}
  P_{A \to AB}  &= s \cdot (\sigma_B)^{a}, \label{eq:bil1}\\
  P_{B \to AB}  &= (1-s)     \cdot (\sigma_A)^{a}, \label{eq:bil2}\\
  P_{AB \to A}  &= (1-s)     \cdot (1-\sigma_B)^{a}, \label{eq:bil3}\\
  P_{AB \to B}  &= s \cdot (1-\sigma_A)^{a}, \label{eq:bil4}
\end{align}
where transitions between A and B are constrained to pass through the bilingual
intermediate state $AB$. The dynamic edge-weighting rule Eq.~\eqref{eq:weight}
carries over naturally: edges between any two non-majority speakers (including
bilinguals) can be reinforced.\cite{mira2010importance, cherniha2020exact,castello2008modelling}

\subsection{Observation from the Results}

Fig.$8$ presents the phase diagrams obtained for the proposed bilingual extension of the language competition model for two representative network sizes, $N=100$ and $N=1000$. For both systems, the language prestige parameter $s$ is varied systematically to investigate its influence on the transition between coexistence and monolingual dominance. The selected prestige values represent the coexistence, critical, and dominance regimes, allowing the evolution of the phase behaviour to be examined across different system sizes.

A common feature observed for both network sizes is the progressive reduction of the coexistence region with increasing language prestige. At relatively low values of $s$, neither language possesses a sufficient social advantage to eliminate the competing language, and the presence of bilingual agents promotes long-term coexistence by acting as an intermediate linguistic state that facilitates communication between the two language communities. As the prestige parameter approaches its critical value, the competition between majority-language expansion and minority-language persistence becomes increasingly balanced, producing a narrow transition region that separates coexistence from dominance. Beyond this threshold, the higher-prestige language spreads rapidly throughout the network, causing the coexistence region to contract significantly and driving the system towards monolingual dominance.

Although the qualitative evolution remains identical for both network sizes, increasing the number of agents leads to a noticeably sharper phase transition. The larger network exhibits a more clearly defined phase boundary and a narrower crossover region, reflecting the reduced influence of finite-size fluctuations. Consequently, the transition between coexistence and dominance becomes considerably more abrupt for $N=1000$ than for $N=100$, while the sequence of dynamical regimes remains unchanged.

These results demonstrate that the introduction of bilingual agents enriches the dynamics of the language competition model by providing an intermediate linguistic state that promotes coexistence over a broader region of the parameter space. However, the stabilising influence of bilingualism is ultimately constrained by language prestige. Once the prestige imbalance exceeds a critical threshold, the advantage of the dominant language overcomes the buffering effect of bilingual interactions, resulting in global language dominance. Overall, the phase behaviour is governed by the combined influence of language prestige, bilingual interactions, and network size, providing a more realistic description of language competition in complex social networks.

\subsection{Expected Phase Behaviour}

Based on the analytic results of \cite{vazquez2010agent}, the introduction of bilingual
agents is expected to shift the critical coexistence volatility downward: a higher
degree of language switching flexibility is required to maintain two-language
coexistence. In our dynamic-weighting framework, the interplay between $\Delta$-reinforcement
and the bilingual buffer state will determine whether bilingualism stabilises
coexistence or, counter-intuitively, accelerates the path to monolingual dominance.

\subsection{Heterogeneous Prestige and Volatility}

A further planned extension implements non-uniform prestige values $s_i$ and
volatility values $a_i$ at the individual agent level, simulating social hierarchies
and varying degrees of linguistic openness.  This heterogeneous agent model is more
realistic and aligns with the sociolinguistic observation that language loyalty varies
substantially across demographic subgroups \cite{milroy1987language,mufwene2003language}.


\section{Conclusions}

We have presented a computational study of language competition on Erd\H{o}s--R\'enyi
random networks incorporating dynamic edge weighting and probabilistic switching.
Our main findings are:

\begin{enumerate}
  \item Dynamic edge weighting ($\Delta$) counteracts the homogenising effect of
        network connectivity ($p$), with a critical threshold $\Delta_c(p)$ above
        which minority language coexistence is sustained.
  \item Phase boundaries identified via SVM shift upward (require higher $\Delta$)
        as $p$ increases, confirming that denser connectivity accelerates language
        shift absent strong minority bonding.
  \item Anomalous minority clusters persist within predicted dominance regions,
        especially near the phase boundary. Their prevalence grows with $N$ in
        absolute terms but shrinks relative to $N$, consistent with a
        thermodynamic-limit argument.
  \item At extended simulation horizons ($T = 750$), anomalous clustering is
        reduced compared to $T = 250$, and larger networks ($N = 500$) show less
        clustering than smaller ones ($N = 50$), confirming the dual role of
        time and network size in resolving transient minority enclaves.
  \item Bilingual extensions and heterogeneous prestige/volatility are natural
        next steps, expected to produce richer phase diagrams and more realistic
        contact scenarios.
\end{enumerate}

Taken together, these results underscore the importance of modelling social tie
strength as a dynamic quantity in language competition, and suggest that policies
aimed at strengthening intra-community bonds among minority speakers may be as
important as prestige-boosting interventions for language maintenance.






\section*{Author Contributions}

\textbf{Somyaranjan Chakra:} Conceptualization, Methodology, Software, Formal analysis, Investigation, Data curation, Visualization.

\textbf{Mohit Anand Madhesia:} Writing -- original draft, Validation, Discussion and interpretation of results.

\textbf{Shradha Mishra:} Supervision, Project administration, Funding acquisition, Writing -- review \& editing.


\begin{center}
\section*{Conflict of interest}
\end{center}

There are no conflicts of interest to declare.

\section*{Acknowledgment}

S C, M A M, S M thank PARAM Shivay for the computational facility under the National Supercomputing Mission, Government of India, at the Indian Institute of Technology (BHU) Varanasi and also IIT (BHU) Varanasi computational facility. S M thanks DST, SERB (INDIA), Project No.: CRG/2021/006945, MTR/2021/000438, and ANRF grant numbered ANRF/ARG/2025/008220/PS for financial support.

\bibliographystyle{apsrev4-2}
\bibliography{citation}

\end{document}